%% file: arxiv_article_v3.tex
\documentclass[12pt,a4paper]{article}

\usepackage[utf8]{inputenc}
\usepackage{graphicx}
\usepackage{caption} 
\usepackage{subcaption} % for subfigures
\usepackage[colorlinks]{hyperref} % for hyperlinks (to section, citations, etc.)
\usepackage{booktabs} % for better tables
\usepackage{tabularx} 
\usepackage[margin=2cm]{geometry}
\usepackage{authblk} % for author affiliations
\usepackage{physics} % for some physics notation (differentials, etc.)
\usepackage[backend=biber,doi=true,style=numeric,sorting=none,maxnames=10]{biblatex} % for bibliography
\usepackage{cleveref} % for better references
\usepackage{algorithm} % for algorithms
\usepackage{algpseudocode} % for pseudocode
\usepackage{gensymb} % for degree symbol
\usepackage{bbm} % for indicator function
\usepackage{multirow} % for multirow in tables

\usepackage{soul}

\title{
Spontaneous gait synchronisation in the wild: exploring the effect of distance and level of interaction}

\author[1]{Adrien Gregorj}
\author[1,2,3]{Zeynep Y\"ucel}
\author[2,4,5]{Francesco Zanlungo}
\author[2,6]{Takayuki Kanda}
\affil[1]{Okayama University, Okayama, Japan}
\affil[2]{ATR International, Kyoto, Japan}
\affil[3]{Ca' Foscari University of Venice, Venice, Italy}
\affil[4]{Osaka International Professional University, Osaka, Japan}
\affil[5]{University of Palermo, Palermo, Italy}
\affil[6]{Kyoto University, Kyoto, Japan}

\date{\today}

\begin{document}
\maketitle

\begin{abstract}
    Gait synchronisation of pedestrians is influenced by a range of factors, including biomechanical properties like leg length, environmental elements such as presence of obstacles  and cognitive aspects like sensory feedback.  Studying gait data collected in ecological contexts offers unique insights into these numerous factors affecting synchronisation which controlled experimental settings may miss.
    This study addresses the challenges in assessment of gait coordination in complex  real-world interactions by leveraging a dataset of uninstructed pedestrian trajectories. The dataset is  recorded in an underground pedestrian street network and annotated for group relation, interaction levels, and instances of physical contact. The main goals of our study is to devise a method to identify gait synchronisation from trajectory data and to provide an in-depth analysis of social factors affecting gait synchronisation in pedestrian groups.
    To that end, we first propose a method to extract gait residuals from pedestrian trajectories, which capture motion of the body caused by gait-induced oscillations. We thereafter apply a suite of analytical techniques spanning both frequency and nonlinear domains. Frequency-based methods, including the Gait Synchronisation Index and Cross Wavelet Coherence, quantify the alignment of oscillatory patterns in gait. Complementary nonlinear measures, such as Lyapunov exponents, determinism, and recurrence quantification metrics, offer deeper insights into the dynamical stability and predictability of coupled gaits.
    We demonstrate  that higher levels of social interaction are associated with increased gait synchronisation, evidenced by smaller variations in stride frequency, relative phase, and higher synchronisation metrics (GSI and CWC). Distances between pedestrians also influence gait synchronisation, with closer distances leading to stronger synchronisation. Nonlinear analyses indicate that dyads with higher levels of social engagement demonstrate more structured and stable gait dynamics. Additionally, triad formation and relative positioning are shown to influence synchronisation, with certain formations (e.g., $\vee$ and $\longleftrightarrow$) showing more stable gait patterns than others (e.g., $\wedge$).
    Overall, our findings suggest that social interactions shape pedestrian gait coordination, with interaction level and distance being key factors.
\end{abstract}

\section{Introduction}

Human walking is a complex and dynamic process that requires the coordination of multiple biomechanical and neuromuscular systems. Central to this process is the gait cycle, which represents the sequence of movements from one foot contact to the next. This cycle encapsulates the intricate interplay between these systems and is fundamental to understanding locomotion.
Gait synchronisation is influenced by a range of factors, including biomechanical properties such as leg length~\cite{nessler2009interpersonal}, physical coupling~\cite{harrison2009horsing}, sensory feedback~\cite{sylos-labini2018human}, and cognitive demands~\cite{zivotofsky2018effects}. Understanding these factors has implications for a wide range of applications, from urban planning to wearable technology and therapeutic interventions.

Studying gait synchronisation in ecological contexts offers unique insights that controlled experimental settings may miss, as real-world interactions often involve a level of complexity absent in laboratory environments. Besides, analysing pedestrian trajectories in natural settings poses challenges, such as noise, variability, and the lack of controlled conditions. This study addresses these challenges by leveraging a dataset of uninstructed pedestrian trajectories, recorded in an underground commercial district. The dataset is annotated with dyadic and triadic relationships, interaction levels (ranging from 0 for no interaction to 3 for strong interaction), and instances of physical contact.

The primary aim of this study is to explore whether, and to what extent, social interaction intensity and physical contact influence gait synchronisation within social groups of pedestrians. To achieve this, we apply a suite of analytical techniques spanning both frequency and nonlinear domains. Frequency-based methods, including the Gait Synchronisation Index~\cite{zivotofsky2012modalityspecific} and Cross Wavelet Coherence~\cite{grinsted2004application}, quantify the alignment of oscillatory patterns in gait. Complementary nonlinear measures, such as Lyapunov exponents, determinism, and recurrence quantification metrics, offer deeper insights into the dynamical stability and predictability of coupled gaits.

\section{Background and literature survey}

\subsection{Gait cycle and its phases}

Key phases and parameters of the gait cycle are often described using the frameworks established by Perry and Whittle~\cite{whittle2014gait, jacquelin2010gait}. A complete gait cycle, or \textbf{stride}, encompasses the time and distance between two successive placements of the same foot, with \textbf{stride length} measuring the distance covered during this cycle. In contrast, a \textbf{step} refers to the time and movement between successive placements of opposite feet, with each stride consisting of two steps. Another critical parameter is the \textbf{step frequency} (or cadence), which denotes the number of steps taken per unit time. Since each gait cycle includes two steps, cadence effectively reflects the frequency of half-cycles in walking, equivalent to twice the stride frequency.

When individuals walk together, their gaits may exhibit a tendency to synchronise, a phenomenon known as \textbf{gait synchronisation}. It involves some extent of \textbf{phase locking}~\cite{pikovsky2001universal}, where the timing difference between steps remains constant. Notably, two prominent forms of synchronisation are \textbf{in-phase}, where the same foot contacts align, and \textbf{anti-phase}, where opposite foot contacts occur simultaneously.

\subsection{Metrics for detecting gait synchronisation in paired walking}

Understanding gait synchronisation between individuals requires metrics that effectively capture interpersonal coordination. A comprehensive review of these metrics is provided by Felsberg et al. (2021)~\cite{felsberg2021spontaneous}, offering a solid foundation for exploring methods used to detect gait synchronisation in paired walking.

One of the earliest approaches to measuring interpersonal synchrony was introduced by Miles et al. (2010)~\cite{miles2010too}, who analysed the distribution of relative phase values between individuals. By comparing the proportion of in-phase values (i.e., relative phase close to $0$) to chance levels, they quantified the degree of synchronisation, laying the groundwork for more sophisticated methods.

Building on traditional techniques, Nessler et al. expanded the analysis of gait synchronisation by incorporating nonlinear methods rooted in chaos theory. Their work utilised tools such as recurrence plot analysis and Lyapunov exponents to capture the complex dynamics underlying interpersonal coordination~\cite{nessler2009interpersonal, nessler2009nonlinear}. These approaches allowed for a deeper understanding of the stability and variability in synchronised walking.

Based on general measure of synchronicity between two oscillators introduced by Tass in~\cite{tass1998detection}, the Gait Synchronisation Index (GSI) has proven to be a reliable measure of interpersonal gait coordination. Zivotofsky et al. demonstrated its utility across diverse contexts~\cite{zivotofsky2012modalityspecific, zivotofsky2018effects}, reporting GSI scores as high as 0.4 when multiple feedback modalities (visual, auditory, tactile, and instructions) were integrated. Further extending its application, Soczawa et al. employed GSI in virtual reality settings to explore gait synchronisation with avatars~\cite{soczawa2020gait, soczawa2020validation}.

More recently, Liu et al. have studied step synchronisation of pedestrians in the wild using visual inspection to identify synchronisation events~\cite{liu2023understanding} and found that social groups synchronised more often than non-related individuals. Inside social groups, dyads were found to have larger proportions of synchronisation than triads. They also conducted controlled experiments and measured the synchronisation of pedestrians walking side by side by measuring the time difference between instants of the feet touching the ground or being lifted. They showed that this could serve as a reliable indicator of belonging to the same group. Finally, they showed that the motion of the head of the pedestrians could be precisely related to the stepping events, and may be used to detect synchronisation events.

\subsection{Frequency domain analysis and wavelets}

When studying gait synchronisation, frequency domain analysis provides a robust framework for understanding the dynamics of gait patterns. By examining the frequency content of gait signals from two individuals, researchers can identify common oscillatory components indicative of synchronisation.

While Fourier analysis is widely used for analysing periodic signals, it has inherent limitations when applied to non-stationary signals with varying frequency composition. Specifically, Fourier analysis produces a single frequency spectrum for the entire signal, which may fail to capture its dynamic nature. Wavelet analysis addresses this limitation by providing a time-frequency representation, enabling the identification of localised oscillatory components. This makes wavelets particularly well-suited for analysing gait signals, which are inherently non-stationary and often exhibit frequency variations over time.

For a foundational understanding of wavelet analysis, Torrence and Compo offer a practical tutorial that explores its basic principles and applications~\cite{torrence1998practical}. Building on this, Issartel et al. demonstrate the utility of wavelets in studying human motor behaviour, providing domain-specific insights relevant to synchronisation analysis~\cite{issartel2006practical}. In the context of pedestrian gait synchronisation, wavelets have been applied to address various research questions. Zivotofsky et al. used wavelet analysis as a supplementary tool to provide evidence of synchronisation, but primarily relied on GSI for their main results~\cite{zivotofsky2018effects}. Bocian et al. used wavelet transform as a way to obtain the instantaneous phase of pedestrian gait before computing a similar metric to measure interpersonal synchronisation between pedestrians walking on a bridge~\cite{bocian2018time}.

Among wavelet-based methods, wavelet coherence has emerged as a particularly powerful metric for detecting synchronisation. Grinsted et al. provide a comprehensive definition and practical applications of wavelet coherence in various fields~\cite{grinsted2004application}. While its potential for capturing time-varying synchronisation dynamics is well-recognised, to the best of our knowledge, cross wavelet coherence has yet to be fully explored in the context of gait synchronisation.

\subsection{Factors influencing gait synchronisation}

Gait synchronisation is influenced by a combination of intrinsic and extrinsic factors that shape how individuals align their movements during walking.

Gait frequency, walking speed, and leg length are intrinsically related, as the length of an individual’s legs influences their natural walking rhythm. Humans tend to select a step length or frequency that minimises metabolic energy consumption at a given walking speed~\cite{zarrugh1974optimization} with the metabolic cost depending on step length and frequency~\cite{cotes1960energy}. Notably, step length is proportional to leg length, with young adults typically exhibiting step lengths of about 75\% of their leg length~\cite{dale2012clinical}. When individuals with differing leg lengths attempt to synchronise, adaptations in stride length or frequency may move them away from their energy-efficient patterns, potentially increasing their metabolic cost. Nessler et al.~\cite{nessler2009interpersonal} found that the leg length difference between two individuals was significantly correlated with the frequency locking and mean frequency difference among participant pairings. The case of female--male pairs would be particularly interesting to investigate, as the average height difference is naturally larger than in same gender pairs~\cite{yucel2020gender}.

Sensory feedback is a another key modulator of gait synchronisation, with different modalities offering varying degrees of effectiveness. Harrison et al.~\cite{harrison2009horsing} demonstrated that the percentage of phase locking increased progressively with the type of feedback provided: 40\% of locking for visual feedback, 63\% for mechanical feedback (physical connection through a foam block), and up to 77\% for combined visuo-mechanical feedback. These results highlight the additive benefits of multimodal feedback in enhancing synchronisation. Sylos-Labini et al.~\cite{sylos-labini2018human} reinforced the importance of tactile cues, showing that spontaneous synchronisation occurred 40\% of the time in 88\% of pairs walking with hand contact. Similarly, Nessler and Gilliland~\cite{nessler2009interpersonal} observed that while step frequency locking was relatively unaffected by sensory manipulations, phase angle locking was significantly influenced. Among the modalities tested, mechanical coupling achieved the highest phase locking at 46.9\%.

Task complexity and cognitive demands also influence synchronisation. Zivotofsky et al.~\cite{zivotofsky2018effects} found that a simple dual task increased spontaneous synchronisation, while a more complex dual task reduced synchronisation, potentially due to cognitive load interfering with the attention required for coordinated movements. Notably, tactile feedback through hand-holding remained effective in enhancing synchronisation across both simple and complex tasks.

The prevalence of spontaneous synchronisation varies considerably across studies, highlighting the nuanced nature of this phenomenon. Hajnal et al.~\cite{hajnal2023how} reported that only 6\% of 498 coded pairs observed in-the-wild exhibited continuous synchronisation, emphasising its rarity in unconstrained conditions. In contrast, Zivotofsky and Hausdorff observed spontaneous synchronisation in 50\% of walking trials~\cite{zivotofsky2007sensory} for pairs of pedestrians walking on a treadmill.
Again in a controlled environment, Zivotofsky et al. found that 36\% of walks exhibited spontaneous synchronisation, with tactile and auditory feedback significantly enhancing coordination~\cite{zivotofsky2012modalityspecific}. These findings suggest that while synchronisation may emerge naturally in some pairs, external cues often play a crucial role in fostering alignment.

\section{Methods}

\subsection{Dataset}
\label{sec:dataset}
\begin{figure}[!htb]
    \centering
    \begin{subfigure}{0.8\textwidth}
        \centering
        \includegraphics[width=\textwidth]{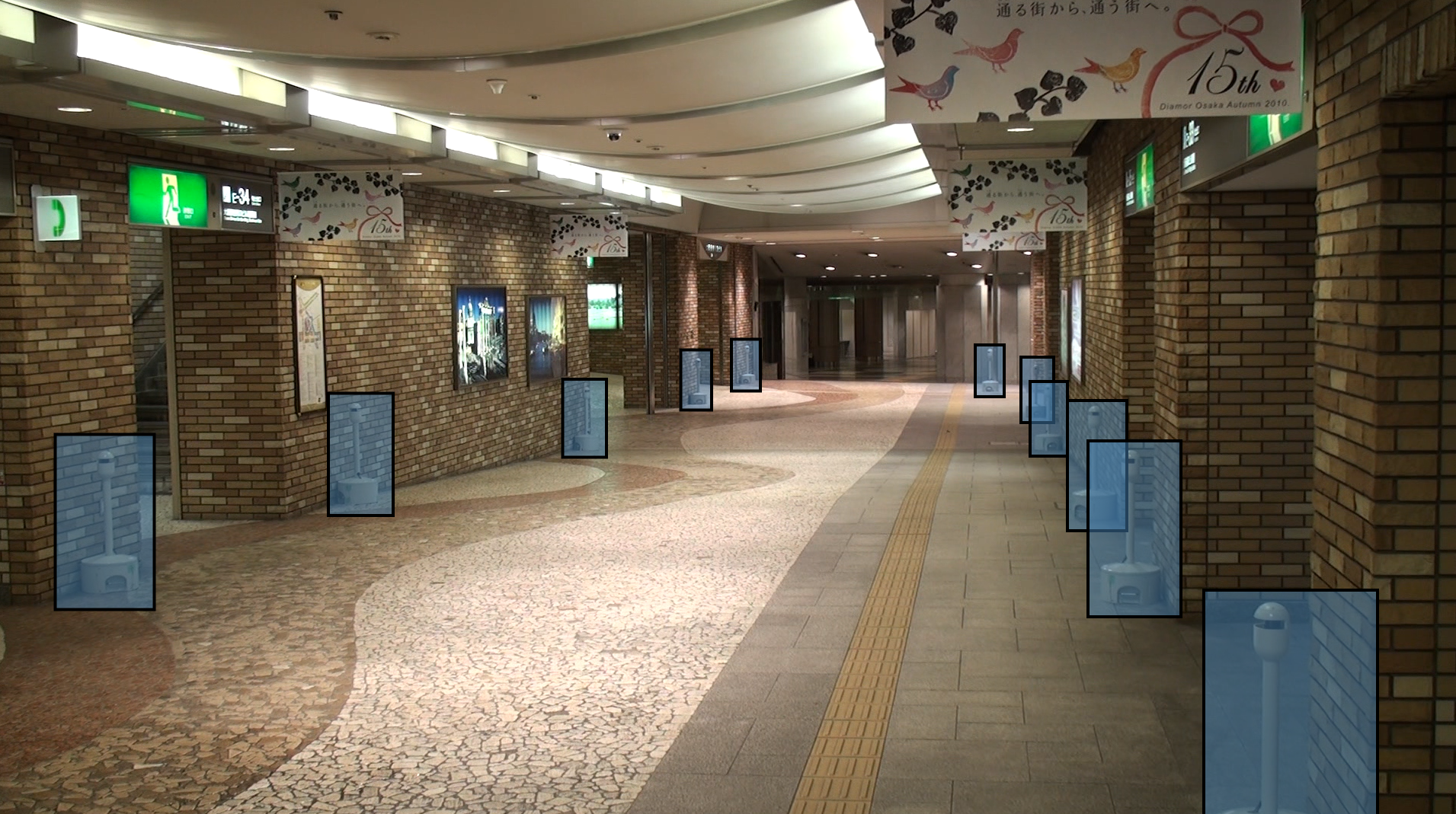}
        \caption{}
        \label{fig:diamor_empty}
    \end{subfigure}
    \begin{subfigure}{0.8\textwidth}
        \centering
        \includegraphics[width=\textwidth]{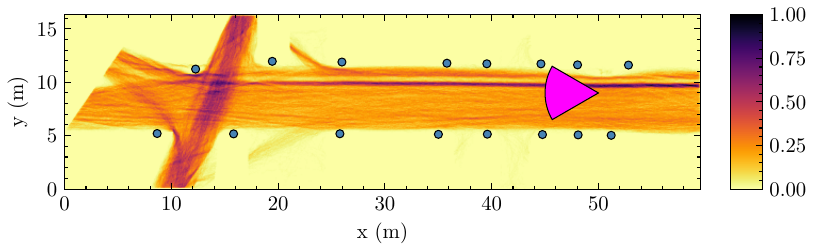}
        \caption{}
        \label{fig:occupancy_grid_diamor}
    \end{subfigure}
    \caption{DIAMOR dataset. (a) Image of the underground pedestrian street network where the DIAMOR dataset was recorded, with tracking sensors marked in blue. (b) Normalised cumulative density map for the DIAMOR dataset (on the first day of recording), created by dividing the area into 10~cm $\times$ 10~cm cells and counting pedestrian presence in each cell. Counts are normalised by the maximum of the grid, with darker areas indicating higher density. The blue dots mark the tracking sensors, and the magenta wedge indicates the camera's field of view.}\label{fig:diamor}
\end{figure}

This study relies on the publicly available DIAMOR dataset~\cite{Diamor_dataset}, which has been extensively used in prior research on pedestrian dynamics and group detection~\cite{glas2014automatic, yucel2018modeling}. The dataset captures pedestrian movement in an underground street network located in Osaka, Japan, a commercial district surrounded by train stations, business hubs, and shopping centres (see~\Cref{fig:diamor_empty}). Experimentation has been reviewed and approved by ATR ethics board with document number 10-502-1.

These data are particularly valuable for capturing pedestrians in their natural environment without direct instructions or experimental constraints, offering insights into realistic locomotion. It can therefore complement controlled studies by providing a broader perspective on pedestrian dynamics.

The dataset is composed of two main components: pedestrian trajectories and video recordings. The pedestrian trajectories were obtained using depth sensors distributed across the underground street network and contain the positions of pedestrians at regular intervals. To reduce noise in the trajectories, we apply a Savitzky--Golay filter to these positions~\cite{savitzky1964smoothing}. The Savitzky--Golay filter is a polynomial smoothing filter that can preserve the shape of the signal while removing noise. We use a window size of $0.25$~s and a polynomial order of $2$ for the filter. The window size is chosen to be big enough to remove measurement noise, but small enough to preserve gait oscillations.

The video recordings were captured using a camera with a field of view covering a portion of the underground street network. The video data were used to annotate the trajectories with information about social groups, interaction levels, and physical contact. A normalised cumulative density map of pedestrian movement is presented in \Cref{fig:occupancy_grid_diamor}. A photograph of the underground street network is shown in \Cref{fig:diamor_empty}, with the sensors used for pedestrian tracking highlighted in blue.

The video recordings were used for identifying two-people and three-people groups and assessing their interaction intensity. The groups were labelled in a two-step process. First, coders determined group membership by observing walking patterns, demographics (e.g., age, gender), and attire. At this stage, they also annotated individuals who appeared to walk independently, without being part of any group (henceforth referred to as \textit{individuals}). Second, they rated the intensity of interaction for identified dyads using a four-level subjective scale (0: no interaction, 1: weak, 2: mild, 3: strong). Coders were not given strict definitions for these levels, but instead viewed three hours of footage involving 2-people groups to develop an intuitive understanding of interaction intensity.

In another annotation step, coders marked instances of physical contact between members of two-people groups, including any form of body contact without constraints on duration.

The agreement between coders for group membership was measured using Cohen's $\kappa$ coefficient, yielding $\kappa = 0.96$, indicative of high reliability~\cite{fleiss2003statistical}. For interaction intensity ratings, Krippendorff's $\alpha$ was used, with a value of $\alpha = 0.67$, which is generally considered acceptable~\cite{krippendorff2004reliability}.

\subsection{Notations}

We start by introducing the notations and definitions used throughout this work. The positions of pedestrians (as obtained by smoothing the original tracking using a 0.25 s Savitzky-Golay filter, as discussed above) are denoted by $\vb{p}(t)$, where $t$ is the time. The velocity of a pedestrian is denoted by $\vb{v}(t)$ and is derived from the positions using a simple forward Euler difference, i.e.\

\begin{equation}
    \vb{v}(t_k) =
    \begin{cases}
        \frac{\vb{p}(t_{k+1}) - \vb{p}(t_k)}{t_{k+1} - t_k} & \text{if } k < N-1 \\
        \vb{v}(t_{k-1})                                     & \text{if } k = N-1
    \end{cases}.
    \label{eq:velocity}
\end{equation}

A \textit{trajectory} $T$ is defined as the sequence of positions $\vb{p}(t_k)$ and velocities $\vb{v}(t_k)$ of the centre  of a pedestrian, sampled at times $t_k$, with $k \in [0, N-1]$ and $N$ being the number of samples. The trajectory is then defined as
\begin{equation}
    T = \left[(\vb{p}(t_0), \vb{v}(t_0)), (\vb{p}(t_1), \vb{v}(t_1)), \ldots, (\vb{p}(t_{N-1}), \vb{v}(t_{N-1}))\right].
\end{equation}

We filter the trajectories to include only those that fall within the range of typical walking speeds commonly observed in public spaces, disregarding anomalies. Drawing on findings from human locomotion studies~\cite{zanlungo2014potential}, we define typical urban walking as having an average velocity between $[0.5, 3]$~m/s. Trajectories outside this range are attributed to activities like standing, running, or potential tracking errors.

\subsection{Social groups}

\subsubsection{Two-person groups: dyads}

We use the term \textit{dyad} to refer to a group of two pedestrians, $i$ and $j$, who share a social relation and walk together toward a common goal, following trajectories $T_i$ and $T_j$.

\input{tables/counts_dyads.tex}

In \Cref{tab:counts_dyads} we show the breakdown of the number of dyads for each level of interaction in the dataset.

\subsubsection{Three-person groups: triads}
\label{sec:triads}

We use the term \textit{triad} to refer to a group of three pedestrians $i$, $j$ and $k$, who share a social relation and walk together toward a common goal, following trajectories $T_i$, $T_j$, and $T_k$.

\begin{figure}
    \centering
    \includegraphics[width=\textwidth]{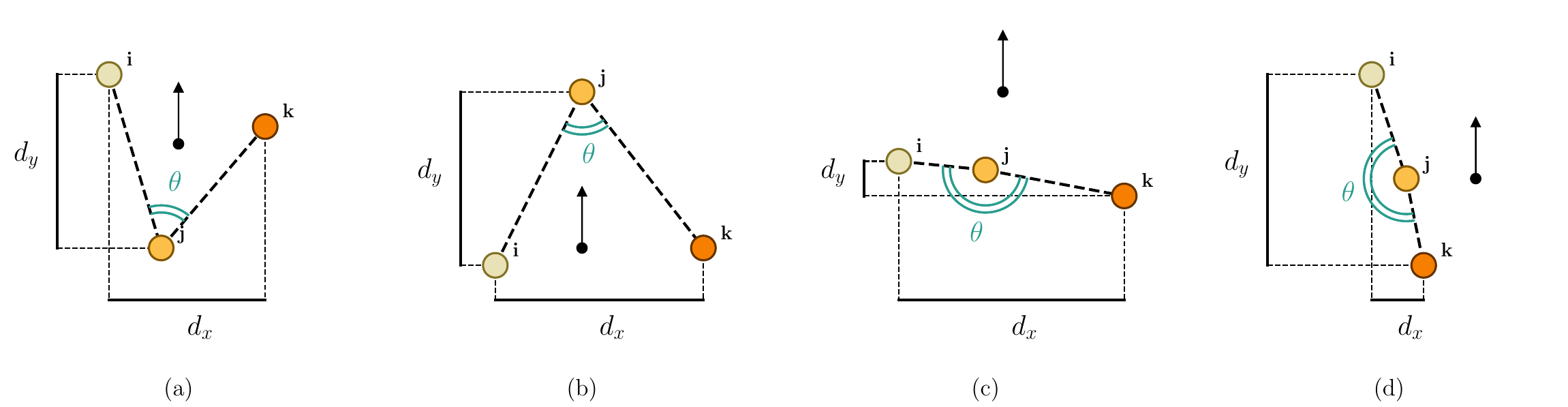}
    \caption{Illustration of the relative positioning of pedestrians in a triad. The four configurations are (a) $\vee$, (b) $\wedge$, (c) $\longleftrightarrow$, and (d) $\updownarrow$.}
    \label{fig:triad_configurations}
\end{figure}

The analysis of gait synchronisation in triads presents additional challenges compared to dyads. In particular, the relative positioning of the pedestrians in the group may be of interest, as it can influence the gait synchronisation between pedestrians (e.g., if one pedestrian is leading the group, the others may adjust their gait to match the leader). Previous studies have classified relative positioning of pedestrians in a triad based on the angles between the vectors connecting them~\cite{costa2010interpersonal}. The authors identified four main configurations: $\vee$, $\wedge$, $\longleftrightarrow$, and $\updownarrow$ (see \Cref{fig:triad_configurations}).

We begin by detailing the method used to perform this classification. Since the triad is mobile, we first perform a change of reference frame to a coordinate system located on the geometric centre of the triad and vertically aligned with the direction of motion of the triad (i.e.\ the $y$-axis is aligned with the velocity vector of the centre of the triad). We consider a geometric definition of the centre of the triad, i.e.\ the average position of the three pedestrians where each pedestrian is weighted equally.

We then compute the average positions for each of the three members across the transformed (translated and rotated) trajectories. We rename the members of the triad such that the member with the lowest average $x$-coordinate (i.e.\ the one on the left, L) is $i$, the member with the highest average $x$-coordinate (i.e.\ the one on the right, R) is $k$, and the remaining member is $j$ (i.e.\ the one in the centre, C). Their average positions vectors are denoted by $\vb{\bar{p}}_i$, $\vb{\bar{p}}_j$, and $\vb{\bar{p}}_k$. We then classify the triad into one of the four configurations by computing the angle $\theta$ between the vector connecting the pedestrians $i$ and $j$ and the vector connecting the pedestrians $j$ and $k$, i.e.\ $\vb{\bar{p}}_j - \vb{\bar{p}}_i$ and $\vb{\bar{p}}_k - \vb{\bar{p}}_j$ respectively. If $\theta$ is between $-160$ and $-20$ degrees, the triad is classified as $\vee$. If $\theta$ is between $20$ and $160$ degrees, the triad is classified as $\wedge$. If $\theta$ is larger than $160$ degrees or smaller than $-160$ degrees, it means that the members are aligned, and we then compute the distance $d_x$ between the pedestrians $i$ and $k$ in the $x$-axis, and the distance $d_y$ between the pedestrians with the highest and lowest $y$-coordinates. If $d_x$ is larger than $d_y$, the pedestrian are walking abreast and the triad is classified as $\longleftrightarrow$. Finally, if $d_x$ is smaller than $d_y$, the pedestrians are following each other and the triad is classified as $\updownarrow$. In that case, the positions horizontally (Left, Right, Centre) are not relevant, and we consider the positions vertically (Forward, Back and Centre).

\begin{figure}
    \centering
    \includegraphics[width=0.6\textwidth]{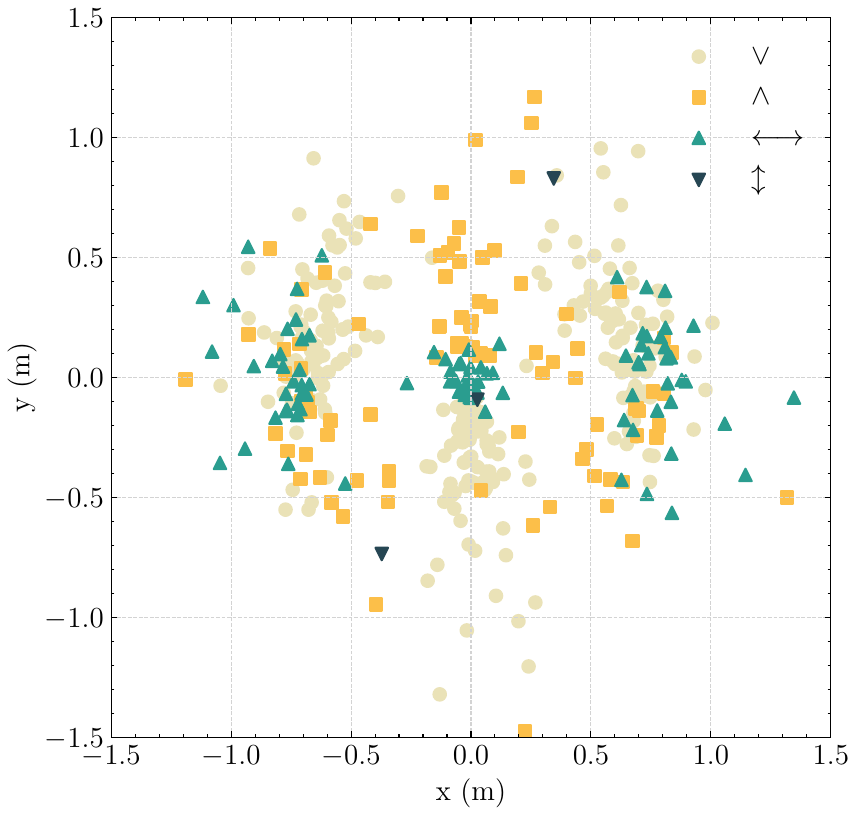}
    \caption{Scatter plot of the average position of the members of triads in the dataset. The four configurations are represented by different colours and symbols.}
    \label{fig:scatter_positions}
\end{figure}

\begin{figure}
    \centering
    \includegraphics[width=\textwidth]{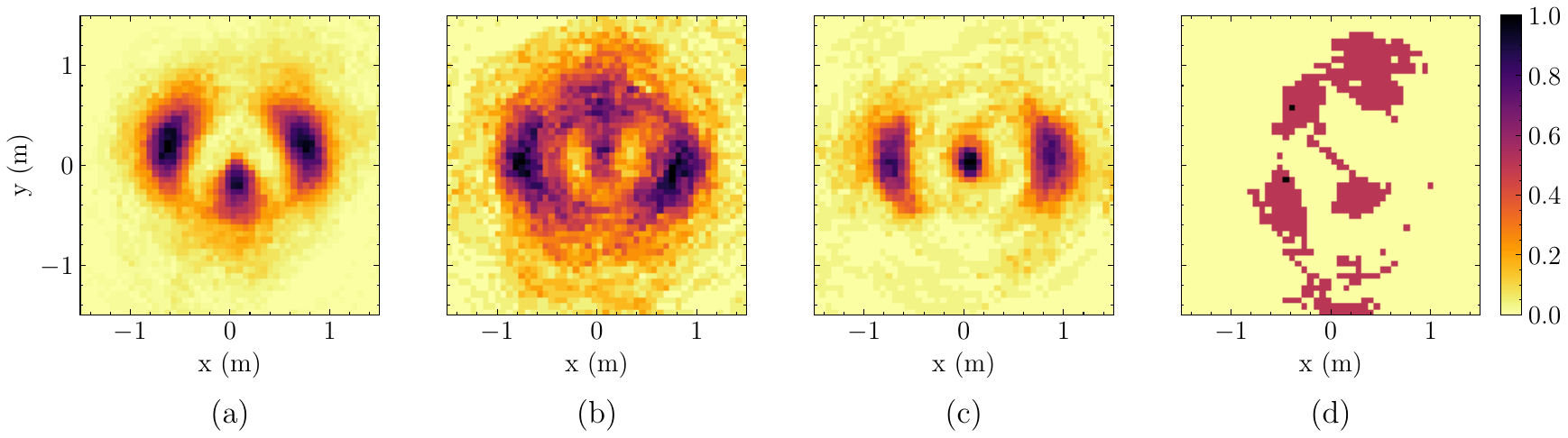}
    \caption{Heatmaps of the instantaneous position of the members of triads in the dataset for each configuration.  (a) $\vee$, (b) $\wedge$, (c) $\longleftrightarrow$ and (d) $\updownarrow$. The colour intensity represents the probability density of the position of the members of the triad.}
    \label{fig:heatmap_positions}
\end{figure}

In \Cref{fig:scatter_positions}, we show the average position of the members for annotated triads in the dataset. The most frequent formation is the $\vee$ configuration, followed by the $\wedge$ and $\longleftrightarrow$ configurations (see \Cref{tab:counts_triads_formations}). The $\updownarrow$ configuration is less common but still occurs in one instance within the dataset. In \Cref{fig:heatmap_positions}, we show heatmaps of the position of the members of the triads in the dataset for each configuration. The colour intensity represents the probability density of the instantaneous position of the members of the triad. It may be noticed that these heatmaps suggest that the $\wedge$ formation is unstable and likely used temporarily for collision avoidance or in crowded environments (see~\cite{zanlungo2015spatial-size}).

The breakdown of the number of triads for each formation is given in \Cref{tab:counts_triads_formations}. We also show the number of pairs of pedestrians in each triad in \Cref{tab:counts_triads_pairs}. Note that the number of pairs is not always the same as the number of dyads, since some pairs might have been discarded during the process (e.g. if we could not compute the gait residuals for one of the pedestrians, etc.).

\input{tables/counts_triads.tex}
\input{tables/counts_triads_pairs.tex}

\subsection{Gait residuals extraction}

In order to retrieve the motion of the centre of the pedestrian caused by gait-induced oscillations, we propose to use the following method. We first compute a smoothed trajectory $\tilde{T}$ by applying another Savitzky--Golay filter to the preprocessed trajectory $T$. For the parameters of the filter, we use a window size of $2$~seconds and a polynomial order of $2$. In contrast to the previous filter used for denoising, the window size is chosen to be larger to flatten the trajectory and remove the oscillations caused by gait.

The gait residuals is the distance between the smoothed trajectory and the original trajectory. More formally, the gait residual $\gamma_k$ at time $t_k$ is defined as the signed distance between the point $\vb{p}(t_k)$ (on trajectory $T$) and its projection onto the line going through the points $\tilde{\vb{p}}(t_{k-1})$ and $\tilde{\vb{p}}(t_{k+1})$ (on trajectory $\tilde{T}$),

\begin{equation}
    \gamma_k =
    \left(\vb{p}(t_k) - \tilde{\vb{p}}(t_k)\right) \cross \frac{\tilde{\vb{p}}(t_{k+1}) - \tilde{\vb{p}}(t_{k-1})}{\norm{\tilde{\vb{p}}(t_{k+1}) - \tilde{\vb{p}}(t_{k-1})}}.
    \label{eq:gait_residual}
\end{equation}

where $\cross$ is the 2D cross product, defined as $\vb{a} \cross \vb{b} = a_x b_y - a_y b_x$ for $\vb{a} = (a_x, a_y)$ and $\vb{b} = (b_x, b_y)$.

\begin{figure}[htb]
    \centering
    \begin{subfigure}[t]{0.45\textwidth}
        \centering
        \includegraphics[width=\textwidth]{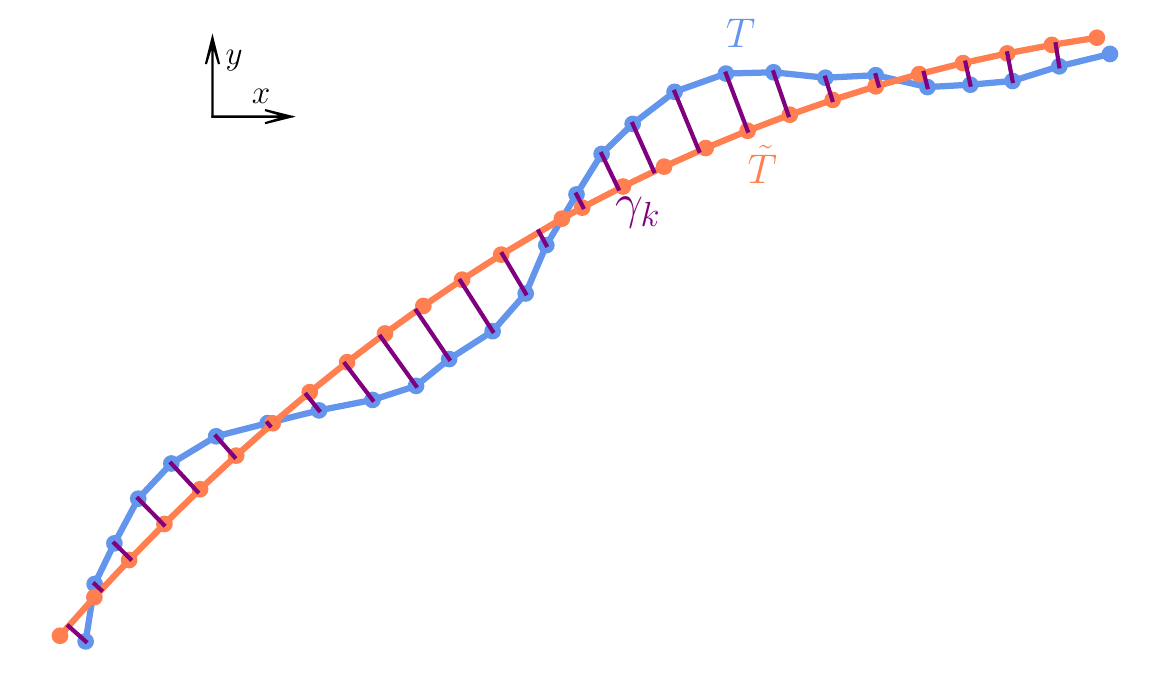}
        \caption{}
    \end{subfigure}
    \begin{subfigure}[t]{0.45\textwidth}
        \centering
        \includegraphics[width=\textwidth]{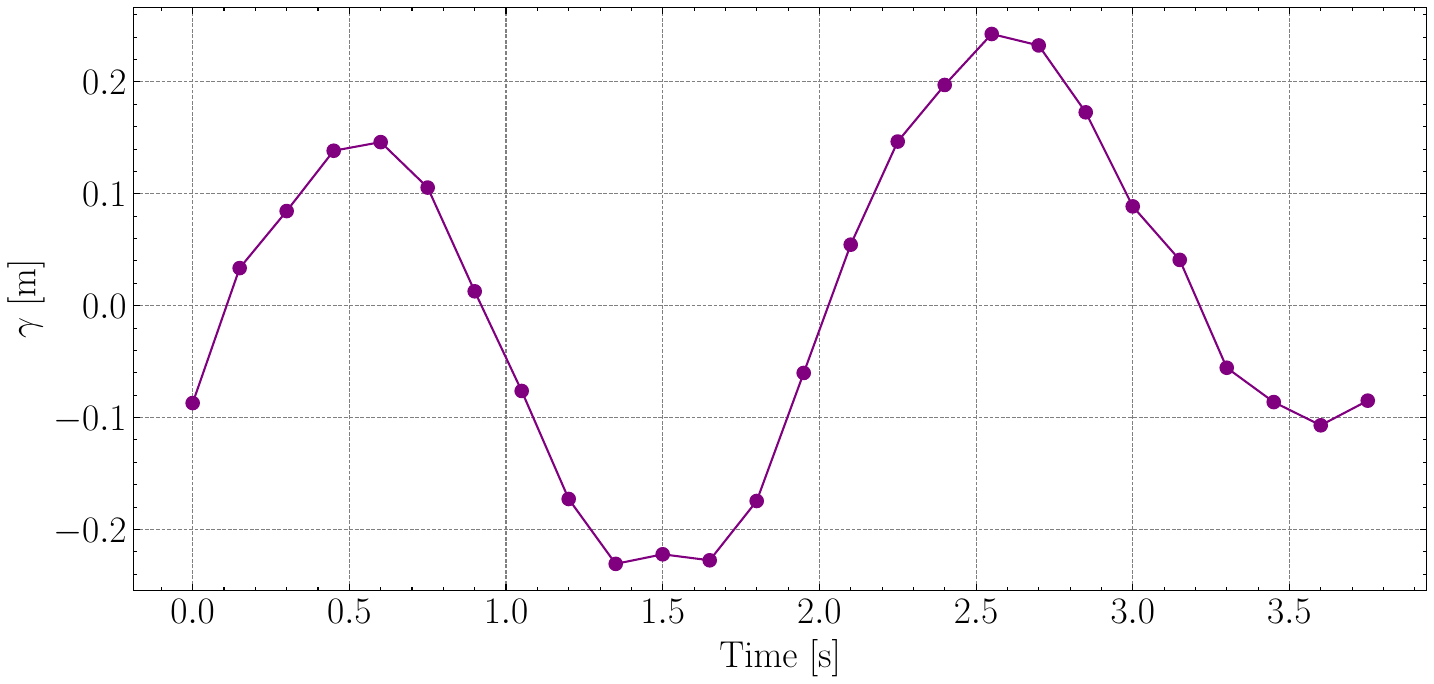}
        \caption{}
    \end{subfigure}
    \caption{Illustration of the gait residual extraction. (a) Hypothetical original trajectory $T$ of a pedestrian in blue and corresponding smooth trajectory $\tilde{T}$ in orange. The gait residuals $\gamma$ are computed as the signed distance from the smoothed trajectory to the original trajectory. (b) Gait residuals obtained for the trajectory in (a).}
    \label{fig:gait_residuals}
\end{figure}

\begin{figure}[htb]
    \centering
    \includegraphics[width=0.7\textwidth]{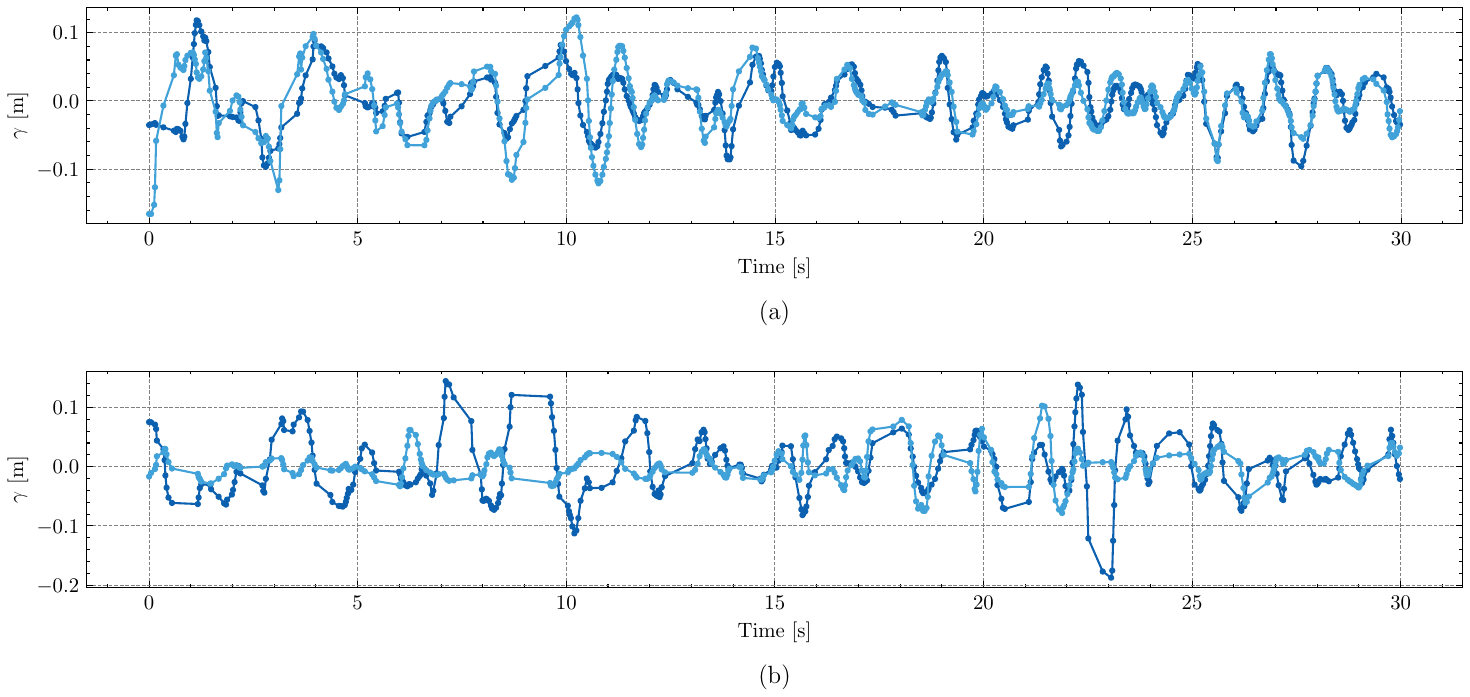}
    \caption{Gait residuals. (a) Gait residuals of two pedestrians walking together in a dyad. (b) Gait residuals of two arbitrary individuals.}
    \label{fig:real_gait_residuals}
\end{figure}

\Cref{fig:gait_residuals} illustrates the gait residual extraction process on an example hypothetical trajectory (the gait induced sway is exaggerated for illustration purposes) and \Cref{fig:real_gait_residuals} shows gait residuals obtained from real pedestrian trajectories in the dataset.

\subsection{Stride frequency estimation}

Computing the stride frequency of a pedestrian from the gait residuals is an important step to verify the effectiveness of the gait residual extraction method and verify that the gait-induced oscillations are correctly captured and consistent with literature values.

To retrieve the stride frequency of a pedestrian from the gait residuals, we use a method similar to the one proposed by Hediyeh et al.~\cite{hediyeh2014pedestrian}, which consists in computing the spectral density of the gait residuals. For a given signal, the spectral density provides information about the frequency content of the signal.

In our case, the spectral density of the gait residuals will reveal the dominant frequency of the gait-induced oscillations. We expect to find a peak in the spectral density at the stride frequency of the pedestrian, and possibly at higher frequencies since the signal will still contain noise caused by the environment, the recording equipment, etc.

We estimated the spectral content of the gait residuals using the \texttt{scipy.signal.periodogram} function, which computes the squared magnitude spectrum. This method applies the Discrete Fourier Transform (DFT) to the signal and returns the spectral power at each frequency. The spectral power $S(f_k)$ at frequency $f_k$ is given by
\begin{equation}
    S(f_k) = \frac{1}{N} \left| \sum_{n=0}^{N-1} \gamma_n e^{-2\pi \mathrm{i} f_k  n \Delta t} \right|^2,
    \label{eq:spectral_density}
\end{equation}
$\mathrm{i}$ denoting the imaginary unit,
for all $k \in [0, N-1]$ where $\gamma_n$ are the gait residuals, and $N$ is the total number of samples. The corresponding frequency values are given by
\begin{equation}
    f_k = \frac{k}{N \Delta t},
    \label{eq:frequency}
\end{equation}
where $\Delta t$ is the sampling interval. The computed spectral power is expressed in $\text{m}^2$, representing the total power at each frequency without normalisation by frequency resolution.

In practice, to prevent capturing non-relevant peaks, we constraint the frequency range to $[0, 4]$~Hz, which encompasses the typical stride frequency of a pedestrian (approximately $1$~Hz). We also ensure that the selected frequency has a power above a certain threshold, which we set to $10^{-4}$~m$^2$ in our analysis.

\begin{figure}
    \centering
    \includegraphics[width=0.7\textwidth]{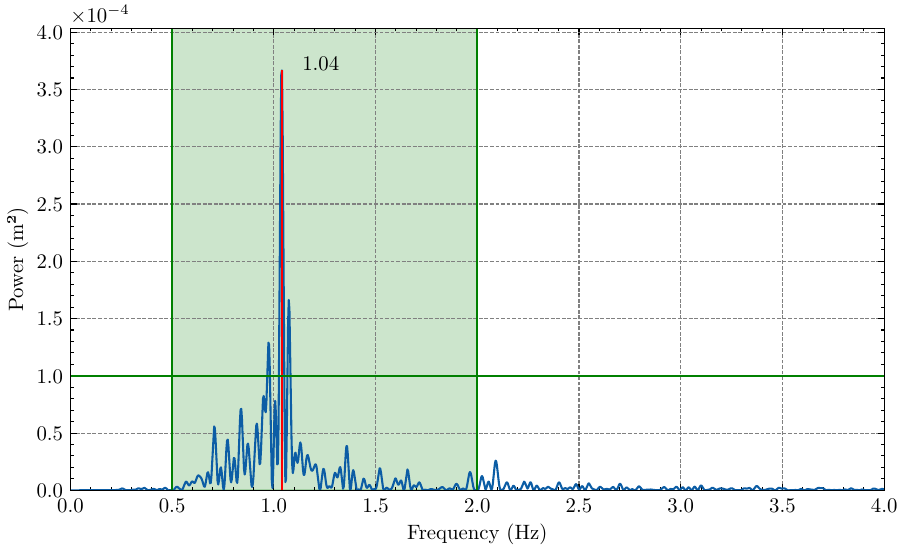}
    \caption{Periodogram of the gait residuals of a pedestrian. The peak at $1.04$~Hz corresponds to the stride frequency of the pedestrian. The green band indicates the range of frequencies of interest ($0.5$ to $2$~Hz) and the horizontal green line indicates the threshold used to determine the stride frequency.}
    \label{fig:periodogram}
\end{figure}

In \Cref{fig:periodogram}, we illustrate the periodogram of the gait residuals of a pedestrian. We observe a peak at $1.04$~Hz, which corresponds to the stride frequency of the pedestrian.

We also compute the stride lengths of the pedestrian by measuring the distance between two consecutive peaks in the gait residuals. The overall stride length is then computed as the average of these distances over the entire trajectory.

\subsection{Gait synchronisation}

\subsubsection{Relative phase}
\label{sec:relative_phase}

The instantaneous phase $\phi$ of a signal $\gamma$ is computed using the Hilbert transform.

The  Hilbert transform is a mathematical operation that takes a real-valued function $\gamma$ and produces a new function, which represents the analytic signal $\gamma_a$.
To obtain the Hilbert transform, the phase of the original signal's Fourier components is shifted by $-\pi/2$ degrees,
\begin{equation}
    H(\gamma)
    =
    \mathcal{F}^{-1}
    \big( \mathcal{F}( \gamma )  2u \big),
    \label{eq:hilbert}
\end{equation}
where $u$ stands for the unit step function.  This allows the construction of a complex analytical signal $\gamma_a$ whose real part is the original signal $\gamma$ and imaginary part is its Hilbert transform $H(\gamma)$,
\begin{equation}
    \begin{aligned}
        \mathcal{R}(\gamma_a)
         & =
        \gamma, \\
        \mathcal{I}(\gamma_a)
         & =
        H(\gamma).
    \end{aligned}
    \label{eq:hilbert_real_img}
\end{equation}
From the analytic signal, we can extract the instantaneous phase $\phi$, which provides valuable information about the signal's frequency content and temporal evolution,
\begin{equation}
    \arg\Big(\gamma_a\Big)
    =
    \phi.
    \label{eq:hilbert_phase}
\end{equation}

The relative phase $\Delta\phi_{ij}$ between two pedestrians $i$ and $j$ is defined as the circular difference between their Hilbert phases $\phi_{i}$  and $\phi_{j}$.

\begin{equation}
    \Delta\phi_{ij}(t_k) = \left[ \left( \phi_{i}(t_k) - \phi_{j}(t_k) + \pi \right) \bmod 2\pi \right] - \pi.
    \label{eq:phase_difference}
\end{equation}

Since we are working with angles, an analysis of the the first and second moments of relative phase necessitates the use of directional statistics~\cite{mardia1999directional}. Specifically, the mean relative phase, denoted as $\overline{\Delta\phi_{ij}}$, is calculated as the circular mean of the instantaneous relative phase over the entire trajectory.

\begin{equation}
    \overline{\Delta\phi_{ij}} = \text{atan2}\left(\frac{1}{N}\sum_{k=0}^{N-1}\sin(\Delta\phi_{ij}(t_k)), \frac{1}{N}\sum_{k=0}^{N-1}\cos(\Delta\phi_{ij}(t_k))\right),
    \label{eq:mean_phase_difference}
\end{equation}

where $\text{atan2}$ is the four-quadrant inverse tangent function.

The circular variance of the relative phase $\sigma_{\Delta\phi}^2$ is computed as

\begin{equation}
    \sigma_{\Delta\phi}^2 = 1 - \left|\frac{1}{N}\sum_{k=0}^{N-1}e^{\mathrm{i}\Delta\phi(t_k)}\right|,
    \label{eq:circular_variance}
\end{equation}
where $\mathrm{i}$ is the imaginary unit.

The circular variance is a measure of the dispersion of the relative phase around the mean relative phase. A circular variance of $0$ indicates that all the relative phases are equal to the mean relative phase, while a circular variance of $1$ indicates that the relative phases are uniformly distributed around the circle.

\subsubsection{Gait Synchronisation Index}

To quantify the synchronisation of gait between two individuals, we first employed a general measure of synchronicity between two oscillators, as introduced by Tass in~\cite{tass1998detection}. This approach was later adapted by Zivotofsky et al.~\cite{zivotofsky2012modalityspecific,zivotofsky2018effects} for the specific context of human gait analysis, where it was termed the Gait Synchronisation Index (GSI). The GSI evaluates the consistency of the empirical relative phase between two pedestrians over time. It is computed by calculating the Shannon entropy of the relative phase distribution, which involves binning the relative phase values into $N_{\text{b}}$ bins and generating a histogram from these values. The GSI is then computed as

\begin{equation}
    \text{GSI} = 1 - \frac{H(\Delta\phi_{ij})}{\log(N_{\text{b}})},
    \label{eq:gsi}
\end{equation}
where $H(\Delta\phi_{ij})$ is the Shannon entropy of the relative phase distribution,
\begin{equation}
    H(\Delta\phi_{ij}) = -\sum_{k=1}^{N_{\text{b}}} p_k \log(p_k),
    \label{eq:shannon_entropy}
\end{equation}
and $p_k$ is the empirically measured probability of the relative phase falling into the $k$-th bin,
\begin{equation}
    p_k = \frac{\sum_{l=0}^{N-1} \mathbbm{I}_{\Delta\phi(t_l) \in I_k}(\Delta\phi_{ij}(t_l))}{N},
    \label{eq:probability}
\end{equation}
where $\mathbbm{I}_{\Delta\phi(t_l) \in I_k}$ denotes the indicator function that takes the value $1$ if the relative phase at time $t_l$ falls into the $k$-th bin $I_k$ and $0$ otherwise.

The GSI ranges between $0$ and $1$, with $1$ indicating perfect synchronisation and $0$ indicating no synchronisation (uniform distribution of the relative phase).

\begin{figure}[htb]
    \centering
    \includegraphics[width=\textwidth]{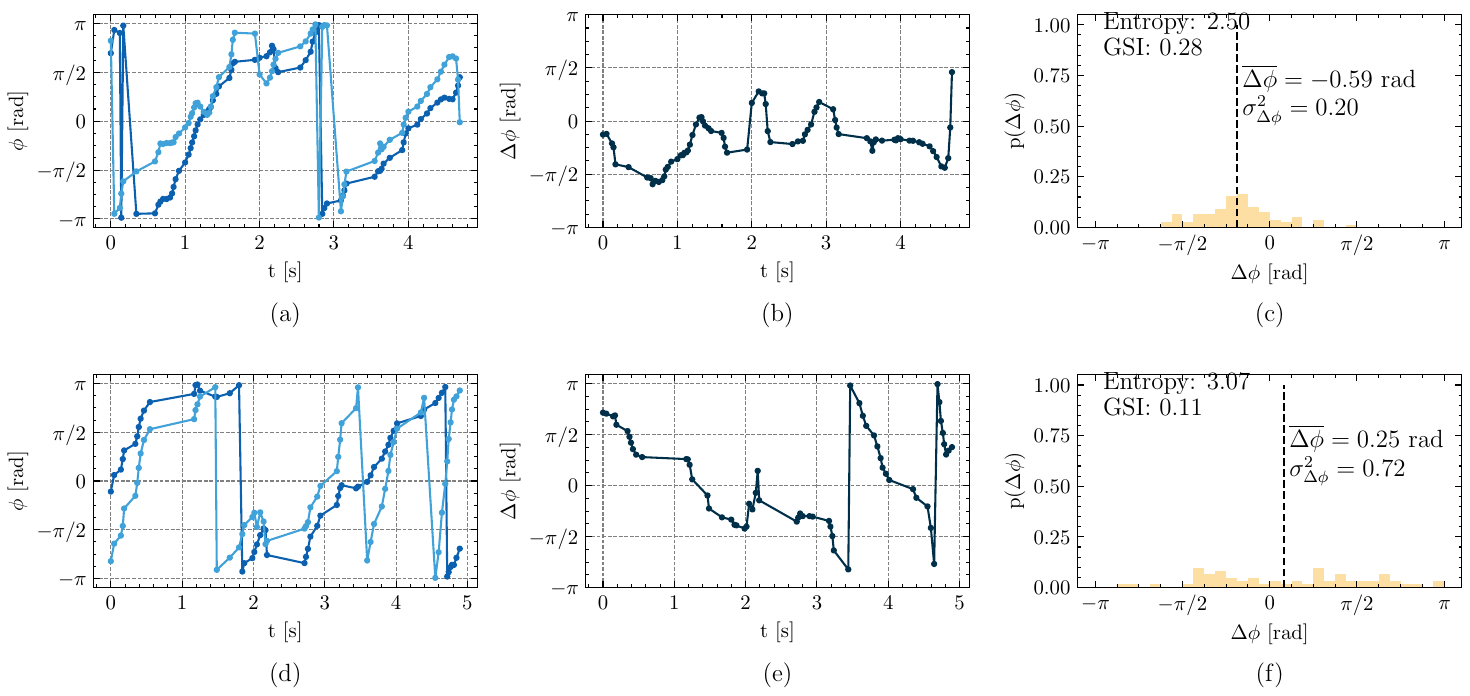}
    \caption{Illustration of the gait synchronisation analysis. (a) Hilbert phase of the gait residuals of two pedestrians in a dyad over a trajectory segment of $5$~seconds. (b) Relative phase between the two pedestrians. (c) Empirical probability density of the relative phase distribution. (d--f) present similar values for two arbitrary individuals.}
    \label{fig:gsi}
\end{figure}

For a given pair of pedestrian trajectories, we compute the GSI and mean relative phase over segments of $5$~seconds and average these values over the entire trajectory. Algorithm \ref{alg:gait_synchronisation} provides the pseudocode for the computation of the GSI and mean relative phase between two pedestrians.

The choice of averaging over segments of $5$~seconds is motivated by the fact that GSI tends to decrease as the length of the segment increases. We posit that this decrease is due to the external factors rather than the intrinsic dynamics of the dyad. Specifically, the dyad members need to avoid  other pedestrians moving in the environment and this collision avoidance behaviour reflects as a  disturbance on their gait synchronisation, limiting the duration that it can be sustained.

%Indeed, when the size of the segment is larger, more values of the relative phase are included in the histogram (in any case, there cannot be less values than for a smaller segment), which will tend to increase the entropy of the distribution and thus decrease the GSI. By averaging over segments of $5$~seconds, we ensure that the GSI is computed over a sufficiently long period to capture the synchronisation between the pedestrians while avoiding the decrease in GSI due to the increase in the segment size.
%In addition, computing the GSI over segments allows us to get results comparable to previous works, such as~\cite{chambers2019pose} where the used videos have a mean duration of $5.1$~seconds.

In \Cref{fig:gsi}-(a--c), we illustrate the gait synchronisation analysis for a dyad, and in \Cref{fig:gsi}-(d--f) for a pair of arbitrary individuals (see \Cref{sec:baseline}). We observe the Hilbert phase of the gait residuals over a segment of $5$~seconds, the relative phase between each pair and the histogram of the relative phase distribution. The corresponding values of the entropy $H$ and the GSI are also shown in \Cref{fig:gsi}-(c,f). We observe that the GSI is higher for the pair of pedestrians walking together in a dyad, indicating a higher level of synchronisation compared to the pair of arbitrary individuals.

\begin{algorithm}
    \caption{Algorithm for computing the GSI and mean relative phase between two pedestrians.}
    \label{alg:gait_synchronisation}
    \begin{algorithmic}[1]
        \Require Trajectories $T_i$ and $T_j$ of pedestrians $i$ and $j$
        \Ensure GSI and mean relative phase
        % loop over split of 5s
        \State Split trajectories $T_i$ and $T_j$ into segments of $5$~seconds
        \State $N \gets \text{number of segments}$
        \For{$k = 0$ to $N-1$}
        \State Compute gait residuals $\gamma_i$ and $\gamma_j$ of segments $T_i[k]$ and $T_j[k]$ using \Cref{eq:gait_residual}
        \State Compute Hilbert phase $\phi_i$ and $\phi_j$ of $\gamma_i$ and $\gamma_j$
        \State Compute relative phase $\Delta\phi_{ij}$ using \Cref{eq:phase_difference}
        \State Compute GSI using \Cref{eq:gsi} and \Cref{eq:shannon_entropy}
        \State Compute mean relative phase using \Cref{eq:mean_phase_difference}
        \EndFor
        \State \Return Average GSI and mean relative phase over all segments
    \end{algorithmic}
\end{algorithm}

\subsection{Wavelet analysis}

Wavelet transform is a powerful tool for analysing non-stationary signals, providing a time-frequency representation that allows us to examine how the frequency content of a signal changes over time. Unlike the Fourier transform, which decomposes a signal into sine and cosine components with infinite duration, thereby losing temporal information, the wavelet transform uses wavelets (short, oscillatory functions with both time and frequency localisation). This makes wavelet transform particularly useful for analysing transient or localised phenomena in signals.

Wavelet analysis involves the decomposition of a signal into wavelets at various scales\footnote{In practice, the scale parameter is related to the bandwidth of the wavelet, with smaller scales involving higher frequencies.} and translations (or time shifts). This decomposition is achieved by convolving the signal with scaled and translated versions of a mother wavelet, a prototype function chosen based on the characteristics of the signal being analysed. The result is a set of coefficients that describe how the signal's frequency content evolves over time.

Mathematically, the continuous wavelet transform (CWT)~\cite{torrence1998practical} of a discrete signal $x$ is defined as

\begin{equation}
    W_x(s, n) = \left(\frac{\delta t}{s}\right)^{\frac{1}{2}} \sum_{m=0}^{N-1} x[m] \psi^*\left(\frac{(m-n)\delta t}{s}\right),
\end{equation}
where $s$ is the scale parameter, $n$ is the translation parameter, $\delta t$ is the sampling interval of $x$, $N$ is the number of samples in the signal, and $\psi^*$ is the complex conjugate of the mother wavelet $\psi$.

The wavelet transform provides a multi-resolution analysis, offering high temporal resolution at small scales and high resolution in the frequency domain at large scales.

Cross wavelet coherence extends the concept of coherence in the frequency domain (using the Fourier transform) to the time-frequency domain~\cite{liu1994wavelet}, allowing for the analysis of the relationship between two signals. It measures the local linear correlation between two signals as a function of both time and scale, revealing how their coherence evolves over time.

The Cross Wavelet Coherence (CWC)~\cite{grinsted2004application} $R_{xy}$ between two signals $x$ and $y$ at scale $s$ and time $n$ is defined as

\begin{equation}
    R^2_{xy}(s, n) = \frac{|S(s^{-1}W_{xy}(s, n))|^2}{S(s^{-1}|W_x(s, n)|^2) \cdot S(s^{-1}|W_y(s, n)|^2)},
\end{equation}

where $W_{x}$ and $W_{y}$ are the wavelet transforms of $x$ and $y$, respectively,
\begin{equation}
    \label{eq:crosswavelet}
    W_{xy} = W_{x}W_{y}^*
\end{equation}
is the cross wavelet transform, and $S$ is a smoothing operator (weighted running average) in both time and scale.

CWC values range between 0 and 1, where values close to 1 indicate strong correlation at a particular time and frequency, and values close to 0 indicate weak or no correlation. This makes cross wavelet coherence particularly useful for detecting and characterising the dynamic interactions between two non-stationary signals across different time scales.

The computation is performed using the \texttt{wct} function from the \texttt{PyCWT} library. We selected the Morlet wavelet as the mother wavelet due to its suitability for analysing oscillatory signals, as it combines a complex wave with a Gaussian envelope. This combination provides good time-frequency localisation, making it ideal for capturing the dynamic interactions in gait patterns. We follow the recommendation of Torrence and Compo~\cite{torrence1998practical} and use a central frequency of 6 (nondimensional) for the wavelet, which balances the trade-off between time and frequency resolution.

\begin{figure}
    \centering
    \centering
    \includegraphics[width=\textwidth]{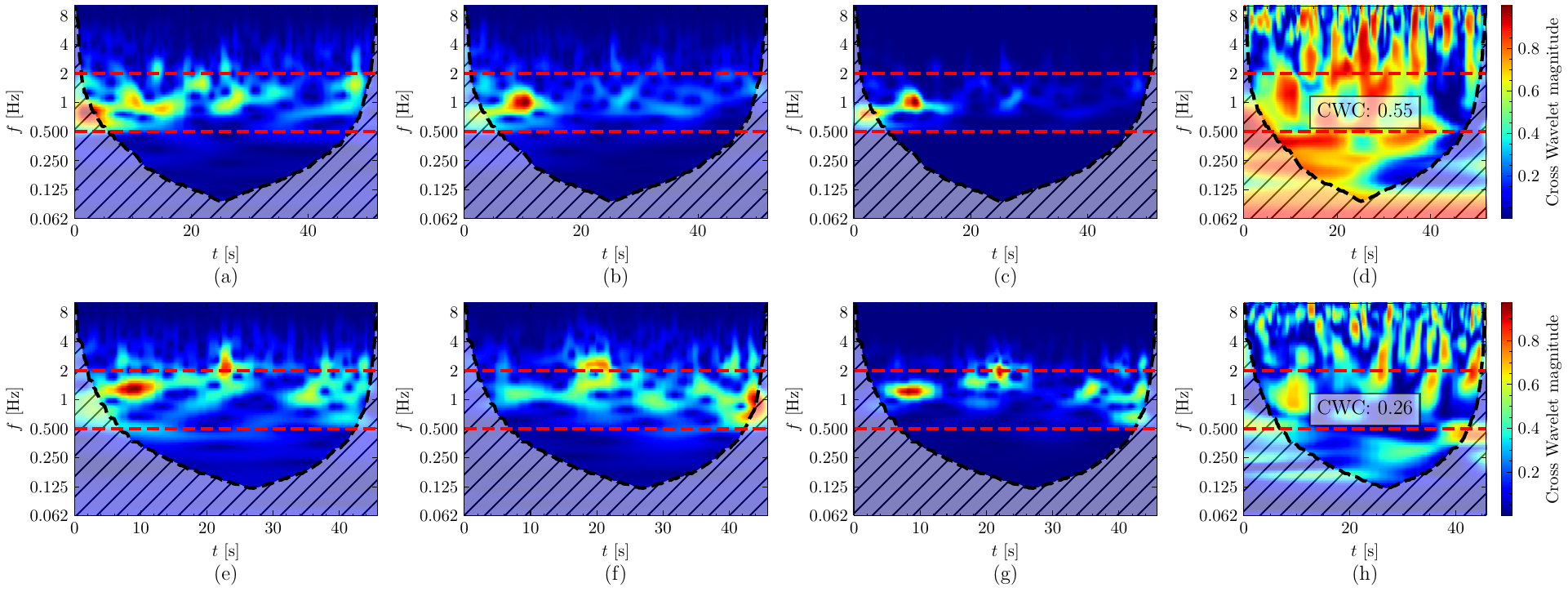}
    \caption{Wavelet analysis of the gait residuals. (a, b) Absolute value of the wavelet transform of the gait residuals of two pedestrians in a dyad. (c) Cross wavelet transform \Cref{eq:crosswavelet} of the gait residuals. (d) CWC between the gait residuals. (e--h) Same as (a--d) for two arbitrary individuals. The red dashed line indicates the band of frequencies of interest ($0.5$ to $2$~Hz). The hatched regions indicate the cone of influence, where edge effects are present.}
    \label{fig:wavelet_analysis}
\end{figure}

In \Cref{fig:wavelet_analysis}-(a)$\sim$(d), we illustrate the wavelet analysis of the gait residuals of two pedestrians in a dyad and in \Cref{fig:wavelet_analysis}-(e)$\sim$(h) we display similar analysis for two arbitrary individuals.
We observe the wavelet transform of the gait residuals for each pedestrian (a, b, e and f), the cross wavelet transform of the gait residuals (c and g), and the CWC between the gait residuals (d and h). We focus on the band of frequencies of interest ($0.5$ to $2$~Hz, see \Cref{fig:periodogram}), which contains the stride frequency of the pedestrians (shown with dashed lines). The hatched regions indicate the cone of influence, where edge effects are present (i.e., where the results are less reliable) due to the zero padding of the wavelet transform~\cite{torrence1998practical}. The CWC reveals the regions of strong correlation between the gait residuals of the two pedestrians, providing insights into the synchronisation of their gait patterns. We see that the absolute value of the wavelet transform of both pedestrians shows high values in the region of the stride frequency (around $1$~Hz), indicating that the gait residuals do contain information about the stride frequency of the pedestrians.

We compute the global CWC of two pedestrians as the average CWC over the entire trajectory~\cite{washburn2014dancers} inside the band of frequencies of interest ($0.5$ to $2$~Hz).
\subsection{Nonlinear analysis}

We employed nonlinear time series analysis to explore the chaotic nature of pedestrian gait patterns~\cite{perc2005dyanmics}. Such methods have been previously used by Nessler et al. to examine gait stability in dyads walking on a treadmill~\cite{nessler2009nonlinear,nessler2013side}. Specifically, we computed the determinism, maximal Lyapunov exponent, and Cross Recurrence Quantification Analysis (CRQA) of the gait residuals.

\subsubsection{Phase space reconstruction}

The first step in nonlinear time series analysis involves reconstructing the phase space of the gait residuals. The reconstructed phase space is a higher-dimensional representation that captures the dynamics of the system. To achieve this, we applied the method of delay embedding~\cite{takens1981detecting}, which reconstructs the phase space by constructing a sequence of vectors composed of time-delayed versions of the original data. Mathematically, this process is expressed as:

\begin{equation}
    \vb{X}(t_k) = \left[\gamma(t_k), \gamma(t_k + \tau),  \gamma(t_k + 2\tau) \ldots, \gamma(t_k + (m-1)\tau)\right],
    \label{eq:phase_space}
\end{equation}

where $\vb{X}(t_k)$ is the reconstructed phase space vector at time $t_k$, $\gamma(t_k)$ is the gait residual at time $t_k$, $\tau$ is the time delay, and $m$ is the embedding dimension.

The fundamental principle of delay embedding is that the dynamics of a complex system, potentially involving many interacting variables, can be inferred from a single observable variable by utilising its time-delayed versions. This approach allows the analysis of the system's behaviour without direct access to all its internal states. A key aspect of the delay embedding method is the selection of the time delay $\tau$ and embedding dimension $m$. The time delay should be chosen to preserve the temporal dynamics of the system, while the embedding dimension must be large enough to capture the system's complexity and underlying structure.

Following the method of Perc~\cite{perc2005dyanmics}, we used the mutual information method to estimate the time delay $\tau$~\cite{fraser1986independent}. Mutual information quantifies the amount of information shared between two variables, indicating how much knowledge of one variable reduces uncertainty about the other. In this study, we computed the mutual information between the gait residuals and their delayed versions to determine the appropriate time delay $\tau$. The value of $\tau$ was selected as the first minimum of the mutual information function, ensuring that it captures the system's intrinsic dynamics.

The mutual information for a time delay $\tau$ is computed by binning the gait residuals $\gamma$ in $N_{\text{bins}}$ bins and computing the probabilities $p_r$ (resp. $p_s$) that $\gamma(t_k)$ falls into the $r$-th (resp. $s$-th) bin. The joint probabilities $p_{rs}$ that $\gamma(t_k)$ and $\gamma(t_k + \tau)$ fall into the $r$-th and $s$-th bins are also computed. The mutual information $I(\tau)$ is then computed as

\begin{equation}
    I(\tau) = \sum_{i=1}^{N_{\text{bins}}} \sum_{j=1}^{N_{\text{bins}}} p_{rs}(\tau) \log\left(\frac{p_{rs}(\tau)}{p_r p_s}\right).
    \label{eq:mutual_information}
\end{equation}

\begin{algorithm}
    \caption{Algorithm for computing the optimal time delay $\tau$ for the phase space reconstruction.}
    \label{alg:time_delay}
    \begin{algorithmic}[1]
        \Require Gait residuals $\gamma$, maximum time delay $\tau_{\text{max}}$, number of bins $N_{\text{bins}}$
        \Ensure Optimal time delay $\tau$
        \State Compute the probabilities $p[r]$ that $\gamma(t_k)$ falls into the $r$-th bin
        \For{$\tau = 1$ to $\tau_{\text{max}}$}
        \State Compute the joint probabilities $p[r][s]$ that $\gamma(t_k)$ and $\gamma(t_k + \tau)$ fall into the $r$-th and $s$-th bins, respectively
        \State Initialise the mutual information $I[\tau] = 0$
        \For{$i = 1$ to $N_{\text{bins}}$}
        \For{$j = 1$ to $N_{\text{bins}}$}
        \State $I[\tau] \gets I[\tau] + p[i][j] \log\left(\frac{p[i][j]}{p[i] p[j]}\right)$
        \EndFor
        \EndFor
        \EndFor
        \State \Return $\tau$ corresponding to the first minimum of $I[\tau]$
    \end{algorithmic}
\end{algorithm}

\Cref{alg:time_delay} provides the pseudocode for computing the optimal time delay $\tau$ for the phase space reconstruction. We chose the maximum time delay $\tau_{\text{max}}$ to be $20$ and the number of bins $N_{\text{bins}}$ to be $10$.

\begin{figure}
    \centering
    \includegraphics[width=0.7\textwidth]{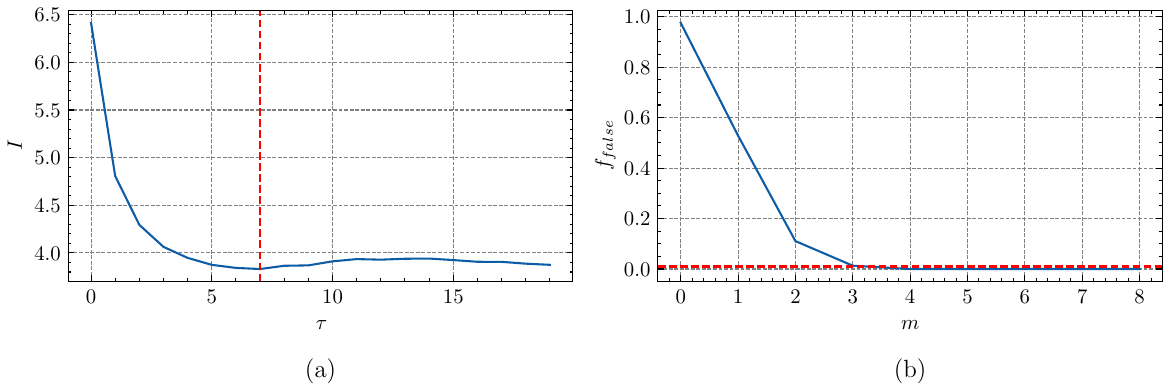}
    \caption{Estimation of the optimal parameters for the phase space reconstruction. (a) Mutual information $I$ as a function of the time delay $\tau$. The optimal time delay is chosen as the first minimum of the mutual information function. (b) Fraction of false nearest neighbours $f_{false}$ as a function of the embedding dimension $m$. The optimal embedding dimension is chosen as the first value for which the fraction of false nearest neighbours is below $1\%$.}
    \label{fig:nonlinear_analysis_parameters}
\end{figure}

In \Cref{fig:nonlinear_analysis_parameters}-(a), we plot the mutual information as a function of the time delay $\tau$ for the gait residuals of one pedestrian in our dataset. The optimal time delay is determined as the first minimum of the mutual information function.

To determine the embedding dimension $m$, we employed the False Nearest Neighbours (FNN) method~\cite{kennel1992determining}. In a reconstructed phase space, a false nearest neighbour of a given point is a point that appears close with a distance smaller than $\epsilon_0$ in a given dimension $m$, but their proximity is merely an artifact of the insufficient embedding dimension so that the points are not close in dimension $m+1$. The FNN method consists in computing the fraction of false nearest neighbours as a function of the embedding dimension $m$, for values of $m$ ranging from $1$ to $m_{\text{max}}$. The embedding dimension $m$ is then chosen as the first value for which the fraction of false nearest neighbours is below a certain threshold, which we set to $1\%$ in our experiments. The pseudo-code for the computation of the optimal embedding dimension $m$ is provided in \Cref{alg:embedding_dimension}.

\begin{algorithm}
    \caption{Algorithm for computing the optimal embedding dimension $m$ for the phase space reconstruction.}
    \label{alg:embedding_dimension}
    \begin{algorithmic}[1]
        \Require Gait residuals $\gamma$, optimum time delay $\tau$, maximum embedding dimension $m_{\text{max}}$, distance threshold $\epsilon_0$, threshold for the fraction of false nearest neighbours $\rho$
        \Ensure Optimal embedding dimension $m$
        \For{$m = 1$ to $m_{\text{max}}$}
        \State Reconstruct two phase spaces $\vb{X}$ and $\vb{X}'$ with delay $\tau$ and embedding dimensions $m$ and $m+1$, respectively
        \State $N_{\text{false}} \gets 0$
        \State $N_{\text{nearest}} \gets 0$ \Comment{Number of points with a nearest neighbour closer than $\epsilon_0$}
        \For{$k = 1$ to $N-1$}
        \State Find the nearest neighbour of $\vb{X}[k]$,  $\vb{X}[k_{\text{nearest}}]$
        \If{$d(\vb{X}[k], \vb{X}[k_{\text{nearest}}]) < \epsilon_0$}
        \State $N_{\text{nearest}} \gets N_{\text{nearest}} + 1$
        \Else
        \State continue \Comment{Skip the point if the nearest neighbour is further than $\epsilon_0$}
        \EndIf
        \If{$d(\vb{X}'[k], \vb{X}'[k_{\text{nearest}}]) > \epsilon_0$}
        \State $N_{\text{false}} \gets N_{\text{false}} + 1$
        \EndIf
        \EndFor
        \State Compute the fraction of false nearest neighbours $f_{\text{false}} = \frac{N_{\text{false}}}{N_{\text{nearest}}}$
        \If{$f_{\text{false}} < \rho$}
        \State \Return $m$
        \EndIf
        \State $m \gets m + 1$
        \EndFor
        \State \Return $m_{\text{max}}$
    \end{algorithmic}
\end{algorithm}

In \Cref{fig:nonlinear_analysis_parameters}-(b), we show the fraction of false nearest neighbours as a function of the embedding dimension $m$ for the gait residuals of a pedestrian in our dataset. We observe that the fraction of false nearest neighbours decreases as the embedding dimension increases.

Note that we ran \Cref{alg:time_delay}  and \Cref{alg:embedding_dimension} over a randomly sampled set of 100 trajectories and found that the average optimal time delay was $6.63$ and the average optimal embedding dimension was $3.62$. We therefore chose $\tau = 7$ and $m = 4$ for the phase space reconstruction of the gait residuals of all pedestrians.
In what follows, we use these parameters to reconstruct the phase space of the gait residuals of all pedestrians.

% \begin{figure}
%     \centering
%     \includegraphics[width=0.7\textwidth]{figures/phase_embedding.pdf}
%     \caption{Phase space reconstruction of the gait residuals of pedestrians. The phase space is reconstructed using the time delay $\tau = 7$ and embedding dimension $m = 4$. First three dimensions of the phase space of two pedestrians for (a) a dyad and (b) two arbitrary individuals.}
%     \label{fig:phase_embedding}
% \end{figure}

% In \Cref{fig:phase_embedding}, we show the phase space reconstruction of the gait residuals of two pedestrians in a dyad and two arbitrary individuals.

\subsubsection{Determinism}

The Kaplan and Glass determinism test~\cite{kaplan1992direct} is a method used to assess whether a time series originates from a deterministic system or a stochastic (random) process. The test examines the structure of reconstructed trajectories in the phase space. To perform the test, the phase space is divided into
$N_{\text{boxes}}$ bins along each dimension, resulting in a total of $(N_{\text{boxes}})^m$ for an $m$-dimensional phase space. The number of boxes should be small enough to ensure that each box captures meaningful dynamical structures rather than just noise, and a too-fine partitioning could lead to sparsely populated boxes. On the other hand, a too-coarse partitioning could lead to the loss of important dynamical information.  We selected $N_{\text{boxes}} = 5$ for our analysis, which provided a good balance, as assessed by visual inspection of the phase space.

For each box, the average normalised direction of all trajectories that pass through the box is computed. If the trajectories are well-aligned, the average of these unit vectors will be close to a unit vector itself, indicating strong determinism. On the other hand, if the trajectories are not aligned and are scattered in various directions, their normalised direction vectors will not align. The average of these unit vectors will result in a vector with a smaller magnitude.

The determinism $D$ is then defined as the average magnitude of the direction vectors across all occupied boxes. The determinism ranges between $0$ and $1$, with $1$ indicating a deterministic system and $0$ indicating a stochastic system. The algorithm for computing the determinism is provided in \Cref{alg:determinism}.

\begin{algorithm}
    \caption{Algorithm for computing the determinism of the gait residuals.}
    \label{alg:determinism}
    \begin{algorithmic}[1]
        \Require Reconstructed phase space $\vb{X}$, number of boxes $N_{\text{boxes}}$, minimum number of trajectories passing through a box $N_{\text{min}}$
        \Ensure Determinism $D$
        \State Divide the phase space into $(N_{\text{boxes}})^m$ boxes
        \State Initialise $D \gets 0$, $c_{\text{box}} \gets 0$
        \For{box in boxes}
        \State Compute the average direction $\vb{d}$ of the trajectories passing through the box
        \If{there is more than $N_{\text{min}}$ trajectories passing through the box}
        \State $D \gets D + \|\vb{d}\|$
        \State $c_{\text{box}} \gets c_{\text{box}} + 1$
        \EndIf
        \EndFor
        \State $D \gets \frac{D}{c_{\text{box}}}$
    \end{algorithmic}
\end{algorithm}

\begin{figure}
    \centering
    \includegraphics[width=0.7\textwidth]{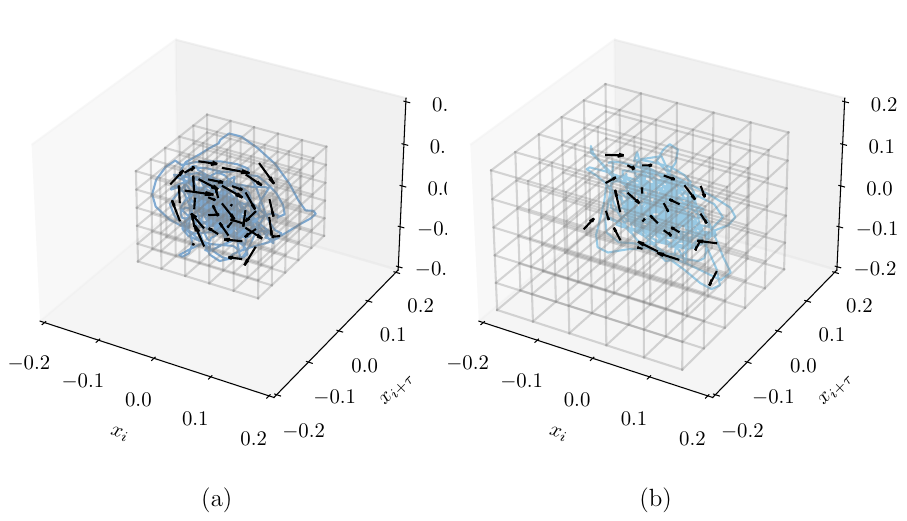}
    \caption{Determinism analysis of the gait residuals of pedestrians. (a) Determinism of a pedestrian from a dyad. (b) Determinism of an arbitrary individual.}
    \label{fig:determinism}
\end{figure}

In \Cref{fig:determinism}, we show examples of the computation of the determinism for the gait residuals of an arbitrary pedestrian in a dyad and a random pedestrian. We plot the three first dimensions of the reconstructed phase space of the gait residuals and the average direction of the trajectories in each box (only for the boxes where the trajectory passes enough times). Although the trajectories are far from being perfectly aligned (given the nature of the gait residuals obtained from real-world data), it appears that the trajectory of the dyad member exhibits larger alignment vectors than the arbitrary individual.

\subsubsection{Maximal Lyapunov exponent}

The maximal Lyapunov exponent $l_{lyap}$ quantifies the rate at which nearby trajectories in a dynamical system's phase space diverge or converge over time.  It is a key measure of the system's sensitivity to initial conditions, which is a hallmark of chaotic behaviour. Systems with positive Lyapunov exponents are characterised by chaotic dynamics, where small differences in initial conditions lead to exponentially diverging trajectories.

To compute the maximal Lyapunov exponent, we use the algorithm proposed by Rosenstein et al.~\cite{rosenstein1993practical}, which estimates the average rate of logarithmic divergence of nearby trajectories. The algorithm begins by selecting a reference point in the reconstructed phase space of the dynamical system and computing the distance between this point and its nearest neighbours. The system is then evolved forward in time, and at each time step, the distance between the reference point and its nearest neighbours is recalculated. This captures how the separation between these trajectories evolves over time.

This process is repeated for multiple reference points and the average divergence for all reference points is computed. The maximal Lyapunov  exponent is then estimated as the slope of the linear fit of the logarithm of the average divergence as a function of time, sometimes referred to as the expansion rate~\cite{perc2005dyanmics}.

\begin{figure}
    \centering
    \includegraphics[width=0.7\textwidth]{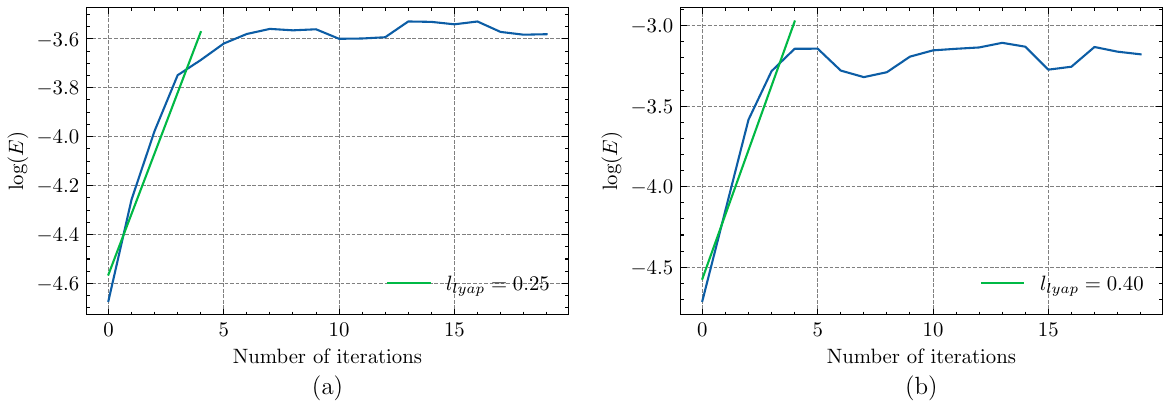}
    \caption{Maximal Lyapunov exponent analysis of the gait residuals of pedestrians. Logarithm of the expansion rate $log(E)$ as a function of the number of iterations for (a) a pedestrian from a dyad and (b) an arbitrary individual. The maximal Lyapunov exponent is computed as the slope of the linear fit.}
    \label{fig:ex_lyapunov}
\end{figure}

The algorithm for computing the maximal  Lyapunov exponent is provided in \Cref{alg:lyapunov} and an example of the computation of the maximal Lyapunov exponent for the gait residuals of a two pedestrians is shown in \Cref{fig:ex_lyapunov}. We observe the logarithm of the expansion rate as a function of the number of iterations for a pedestrian from a dyad and an arbitrary individual. The maximal Lyapunov exponent is computed as the slope of the linear fit over the first five iterations.

\begin{algorithm}
    \caption{Algorithm for computing the maximal Lyapunov exponent of the gait residuals.}
    \label{alg:lyapunov}
    \begin{algorithmic}[1]
        \Require Reconstructed phase space $\vb{X}$, number of reference points $N_{\text{points}}$, number of iterations $N_{\text{iterations}}$, minimum number of nearest neighbours $N_{\text{neigh}}$, distance threshold $\epsilon_1$
        \Ensure Maximal Lyapunov exponent $l_{lyap}$
        % expansion rate is array of size N_steps where each value is the average distance between the reference point and its nearest neighbours
        \State Initialise expansion rate array $E$ of size $N_{\text{iterations}}$
        \State Initialise $n \gets 0$
        \While{$n < N_{\text{points}}$}
        \State Randomly select a reference point $\vb{X}[k]$ in the reconstructed phase space
        \If{$\vb{X}[k]$ has less than $N_{\text{neigh}}$ nearest neighbours closer than $\epsilon_1$}
        \State \textbf{continue}
        \EndIf
        \For{$s = 0$ to $N_{\text{iterations}}$}
        \State Evolve the system forward in time by $s$ iterations
        \State Compute the average distance $d_{n,s}$ between $\vb{X}[k+s]$ and its $N_{\text{neigh}}$ nearest neighbours after $s$ iterations
        \State $E[s] \gets E[s] + d_{n,s}$
        \EndFor
        \State $n \gets n + 1$
        \EndWhile
        \State Compute the maximal Lyapunov  exponent $l_{lyap}$ as the slope of the linear fit of $\log(E)$ as a function of the number of iterations

    \end{algorithmic}
\end{algorithm}

\subsubsection{Cross Recurrence Analysis}

Cross Recurrence Analysis (CRA)~\cite{zbilut1998recurrence} is a method for investigating the relationship between two time series by identifying moments when their dynamics exhibit similar patterns. Unlike traditional recurrence analysis~\cite{webber1994dynamical}, which examines recurring patterns within a single time series, CRA extends this concept to analyse the interactions and dependencies between two distinct systems. CRA has been used across a wide range of scenarios including joint cooperative motor tasks~\cite{ramenzoni2011joint,howell2020youth}, social motor tasks (e.g. conversation, games)~\cite{tolston2014cross-recurrence,paxton2017interpersonal}, and cognitive tasks~\cite{pellecchia2005concurrent}. However to the best of our knowledge, this is the first study applying it on spontaneous gait synchronisation in ecological settings.

In CRA, phase space reconstruction is applied to both time series to detect moments when the trajectories of the two systems approach each other. The proximity between two points in the reconstructed phase space is quantified using Euclidean distance. When the distance between two points falls below a specified threshold $\epsilon_2$, they are considered to be in a state of recurrence. A recurrence matrix is then created, where points in a state of recurrence are marked with a value of $1$ and all other points with a value of $0$.

From the recurrence matrix, three key metrics are computed as described in \cite{shockley2002cross}.
\begin{itemize}
    \item \textbf{Percentage of recurrence ($\%\text{REC}$)}: It represents the ratio of recurrence points to the total number of points. This metric quantifies the proportion of time during which the two systems exhibit similar behaviour.
    \item \textbf{Percentage of determinism ($\%\text{DET}$)}: It quantifies the proportion of recurrent points that form diagonal lines in the recurrence matrix. These diagonal lines indicate intervals during which both systems show similar behaviour. In purely random or stochastic systems, recurrences occur sporadically and are uncorrelated, leading to isolated recurrence points. On the other hand, deterministic systems exhibit more structured and predictable recurrences, resulting in diagonal lines in the recurrence matrix.
    \item \textbf{Maximum line length ($\text{MAXLINE}$)}: The length of the longest diagonal line, representing the longest period of recurrence in the two systems. Longer diagonal lines suggest more sustained and consistent interactions or similarities between the systems.
\end{itemize}

\begin{figure}
    \centering
    \includegraphics[width=0.7\textwidth]{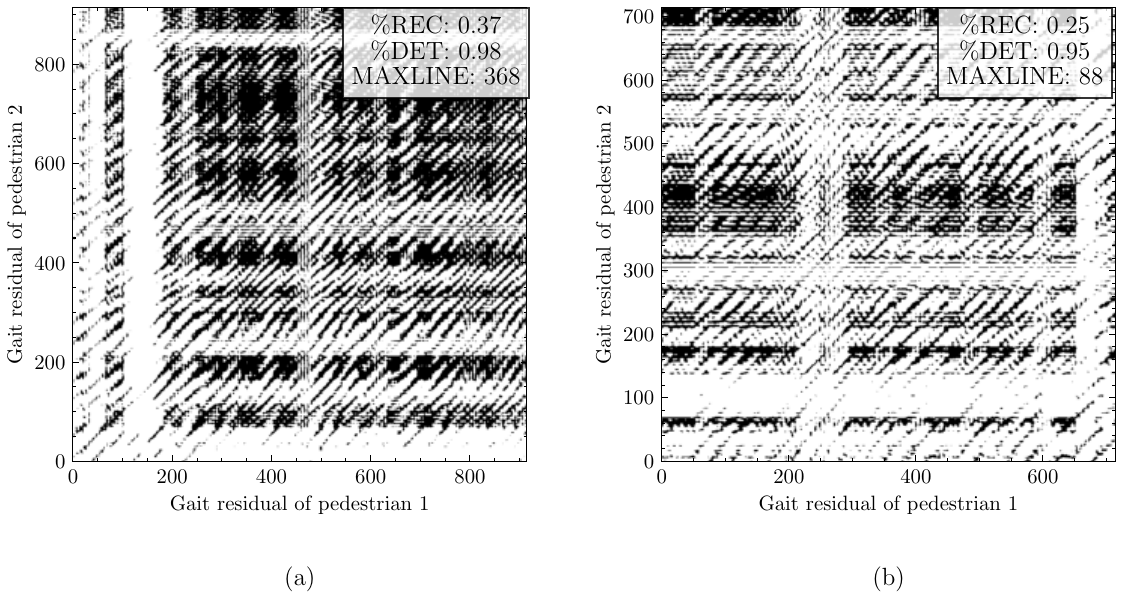}
    \caption{Cross Recurrence Analysis of the gait residuals of pedestrians. (a) Recurrence matrix of the gait residuals of two pedestrians in a dyad. (b) Recurrence matrix of the gait residuals of two arbitrary individuals. The $x$-axis and $y$-axis represent the index of time instants in the trajectories of the two pedestrians, where each point indicates a comparison between two specific moments in the trajectories of the individuals. The matrix is binary, with black points indicating time instants where the gait residuals are sufficiently close. Values for the percentage of recurrence $\%\text{REC}$, percentage of determinism $\%\text{DET}$, and maximum line length $\text{MAXLINE}$ are also shown.}
    \label{fig:cross_recurrence}
\end{figure}

% table for all the parameters of the nonlinear analysis

In \Cref{fig:cross_recurrence}, we show the recurrence matrix of the gait residuals of two pedestrians in a dyad and two arbitrary individuals. All the parameters used for the nonlinear analysis of the gait residuals are summarised in \Cref{tab:nonlinear_parameters}.

\begin{table}
    \centering
    \caption{Parameters used for the nonlinear analysis of the gait residuals.}
    \begin{tabular}{ll}
        \toprule
        Parameter                                                             & Value  \\
        \midrule
        Maximum time delay $\tau_{\text{max}}$                                & $20$   \\
        Number of bins $N_{\text{bins}}$                                      & $30$   \\
        Maximum embedding dimension $m_{\text{max}}$                          & $10$   \\
        Threshold for the fraction of false nearest neighbours $\rho$         & $1\%$  \\
        Distance threshold $\epsilon_0$                                       & $0.07$ \\
        \midrule
        Time delay $\tau$                                                     & $7$    \\
        Embedding dimension $m$                                               & $4$    \\
        \midrule
        Number of boxes $N_{\text{boxes}}$                                    & $5$    \\
        Minimum number of trajectories passing through a box $N_{\text{min}}$ & $3$    \\
        \midrule
        Number of points $N_{\text{points}}$                                  & $100$  \\
        Number of iterations $N_{\text{iterations}}$                          & $5$    \\
        Minimum number of nearest neighbours $N_{\text{neigh}}$               & $1$    \\
        Distance threshold $\epsilon_1$                                       & $0.07$ \\
        \midrule Distance threshold $\epsilon_2$                              & $0.07$ \\
        \bottomrule
    \end{tabular}
    \label{tab:nonlinear_parameters}
\end{table}

\subsection{Baseline}
\label{sec:baseline}

To compare with the gait synchronisation metrics between pedestrians introduced in the previous sections, we also compute these metrics for two baseline scenarios: randomly paired pedestrians and pairs of pedestrians walking close to each other but without  being part of the same dyad.

\subsubsection{Baseline $B_r$ with randomly paired pedestrians}

The first baseline scenario involves randomly selecting two pedestrians labelled as individuals  from the dataset and computing the gait synchronisation metrics between them.

The algorithm for generating the baseline is given in \Cref{alg:baseline}. We randomly select two pedestrian and compute the gait synchronisation metrics between them. We repeat this process until we have computed the metrics for $1000$ pairs of pedestrians.

\begin{algorithm}
    \caption{Algorithm for generating the baseline $B_r$ for gait synchronisation metrics.}
    \label{alg:baseline}
    \begin{algorithmic}[1]
        \Require Set of pedestrian trajectories $\mathcal{P}$
        \Ensure Gait synchronisation metrics $B_r$
        \State $N \gets 1000$
        \State $k \gets 0$
        \State $B_r \gets \{\}$
        \State $\mathcal{U} \gets \{\}$
        \While{$k < N$}
        \State Select $A, B \in \mathcal{P}$ \Comment{Randomly select individuals $A$ and $B$}
        \If{$(A, B) \in \mathcal{U}$}
        \State \textbf{continue}
        \EndIf
        \State $l \gets \text{min}(\text{length}(A), \text{length}(B))$
        \State $A \gets A[:l]$, $B \gets B[:l]$ \Comment{Truncate trajectories to the same length}
        \State $B_r \gets B_r \cup \text{compute\_metrics}(A, B)$
        \State $k \gets k + 1$
        \State $\mathcal{U} \gets \mathcal{U} \cup \{(A, B)\}$
        \EndWhile
        \State \Return $B_r$
    \end{algorithmic}
\end{algorithm}

\subsubsection{Baseline $B_c$ with pedestrians walking close to each other}

The second baseline scenario involves selecting pairs of pedestrians who are walking close to each other  and who are both labelled as individuals. We require the distance between the two pedestrians to be less than $2$~m for at least $10$~s to consider them as a valid baseline pair. This criterion ensures that the pedestrians are walking close to each other for a sufficient time to capture any potential gait synchronisation.

We then compute the gait synchronisation metrics between these pairs.

\section{Results}

\subsection{Gait parameters analysis}

\begin{figure}
    \centering
    \includegraphics[width=\textwidth]{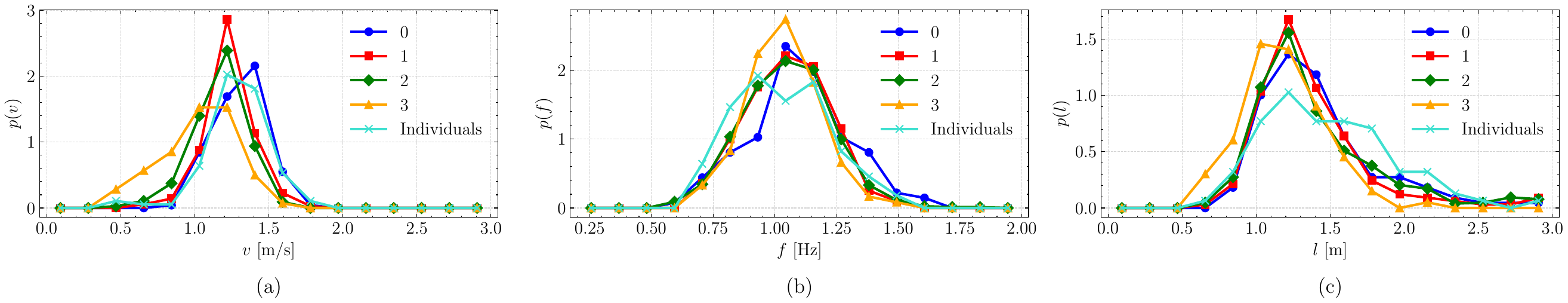}
    \caption{Distribution of gait parameters of dyad members with different levels of interaction.  Probability density functions of (a) velocity $v$, (b) stride frequency $f$, and (c) stride length $l$.}
    \label{fig:gait_stats}
\end{figure}

In \Cref{fig:gait_stats}, we consider distributions of  velocity, stride frequency  and stride length for dyads. The velocity distribution aligns with the expected walking speed of pedestrians, exhibiting a mean value of $1.18$~m/s, and with 90\% of the data falling within the range of $0.84$ to $1.47$~m/s. These values are similar to those reported in previous studies~\cite{zanlungo2015spatial-size,costa2010interpersonal}.

The mean stride frequency for dyads is $1.05$~Hz, with a standard deviation of $0.18$~Hz (see \Cref{tab:gait_stats}). The stride length has a mean value of $1.41$~m and a standard deviation of $0.52$~m. These values are consistent with the typical stride frequency and stride length of pedestrians reported in the literature~\cite{kirtley1985influence,saunier2011estimation,hediyeh2014pedestrian}. The alignment of this distribution with established norms reinforces the reliability of our approach in accurately capturing pedestrian gait patterns from the dataset, thereby strengthening the validity of the following analyses.

We also perform an analysis of variance (Kruskal-Wallis test) to investigate the effect of the level of interaction on the gait parameters of dyad members. The results are presented in \Cref{tab:gait_stats}. We observe a significant effect of the level of interaction on velocity, with the mean velocity decreasing as the level of interaction increases ($p < 10^{-4}$). On the other hand, the stride frequency does not show a significant difference between the levels of interaction ($p = 1.81 \times 10^{-1}$). This means that the decrease in velocity observed for higher levels of interaction is likely due to a decrease in stride length, as the stride frequency remains relatively stable. This is further confirmed by the stride length analysis, where we observe a significant decrease in stride length with increasing interaction ($p < 10^{-4}$).

To compare the gait parameters of dyads with those of individuals, we also show the distribution of gait parameters for individuals in the dataset in \Cref{fig:gait_stats} and the corresponding statistics (mean and standard deviation) in \Cref{tab:gait_stats}. It is important to note that stride length is not derived from velocity and stride frequency. The calculations for each parameter are performed independently, ensuring that stride length is not a direct function of velocity and stride frequency. Namely, we measured the distance between two consecutive peaks in the gait residuals and computed the stride length as the average of these distances over the entire trajectory.

We observe that the mean stride frequency of individuals is similar to that of dyads, but that the former have a significantly higher mean velocity and stride length.

\input{tables/gait_stats.tex}

\begin{figure}
    \centering
    \includegraphics[width=\textwidth]{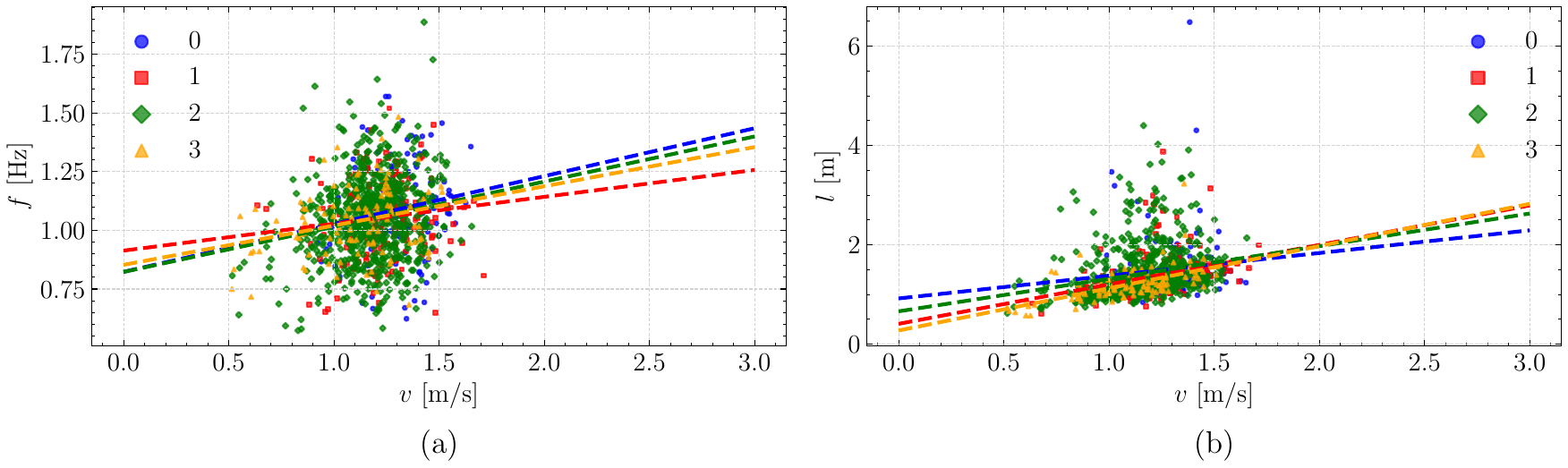}
    \caption{Correlation between gait parameters of dyad members. (a) Velocity $v$ vs stride frequency $f$. (b) Velocity $v$ vs stride length $l$.}
    \label{fig:gait_stats_correlation}
\end{figure}

In \Cref{fig:gait_stats_correlation}, we show scatter plots of the gait parameters of dyad members.
The corresponding Pearson correlation coefficients are presented in \Cref{tab:pearson_correlation}. We observe a slight positive correlation between stride frequency and velocity, with Pearson correlation coefficients ranging from $0.12$ to $0.27$ depending on the level of interaction. We also observe a stronger positive correlation between stride length and velocity, with Pearson correlation coefficients ranging from $0.11$ to $0.58$. These results tend to show that pedestrians who walk faster tend to have longer strides rather than a higher stride frequency~\cite{sekiya1997optimal,sekiya1998reproducibility}.

\input{tables/pearson_correlation.tex}

\subsection{Gait synchronisation analysis}

In this section, we present the results of the gait synchronisation analysis conducted on the pedestrian dataset.

\subsubsection{Effect of the level of interaction in dyads}

\begin{figure}
    \centering
    \includegraphics[width=\textwidth]{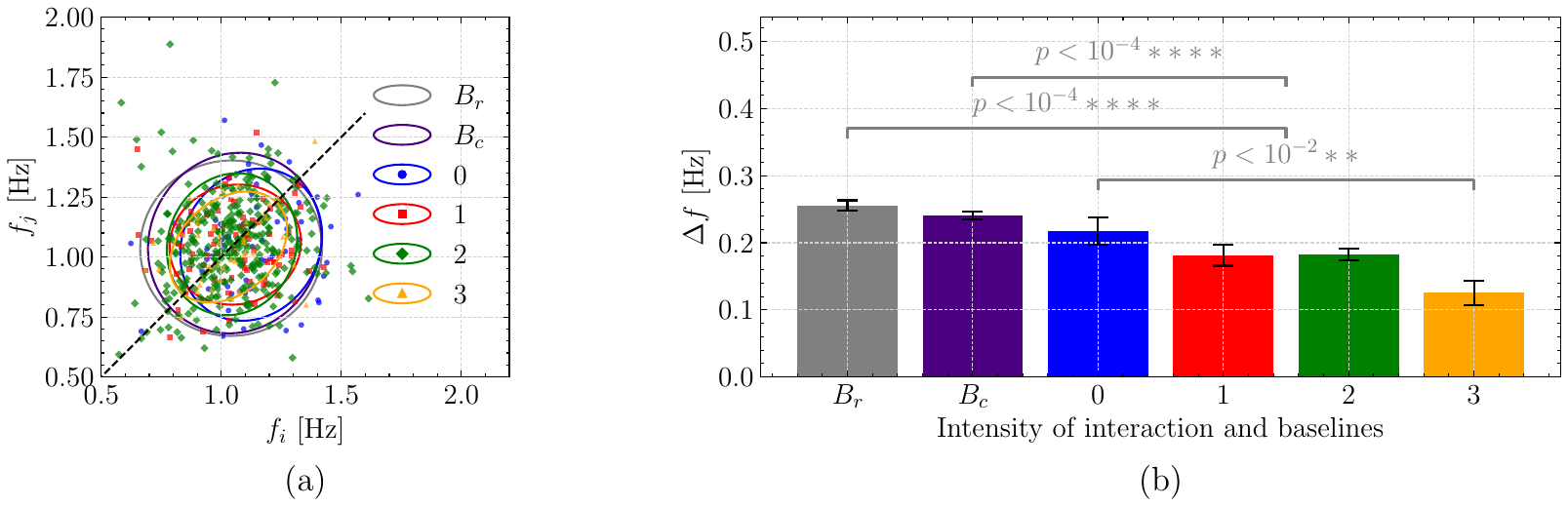}
    \caption{Stride frequency analysis of pedestrian dyads. (a) Scatter plot of the stride frequencies of member $i$ and member $j$ of dyads. The ellipses represent the $95\%$ confidence interval. The purple ellipses corresponds to the baselines $B_r$ and $B_c$, but the individual points are not shown for clarity. (b) Box plot of the difference in stride frequency $\Delta f$ between baseline pairs of $B_r$ and $B_c$ as well as members of dyads.}
    \label{fig:frequency_analysis}
\end{figure}

\input{tables/delta_f.tex}
\input{tables/dunn_delta_f.tex}
\input{tables/ssmd_delta_f.tex}

We start by investigating the effect of the level of interaction on the stride frequency of pedestrians in dyads. In  \Cref{fig:frequency_analysis}-(b), we present a scatter plot of the stride frequencies of member $i$ and member $j$ of dyads. To illustrate the spread of the data, we compute the $95\%$ confidence interval and plot it as an ellipse. We see that higher levels of interaction tend to have a smaller spread and be closer to the $y=x$ line, indicating that pedestrians in dyads tend to have similar stride frequencies. We also observe that the baselines $B_r$ and $B_c$ have a larger spread.  \Cref{fig:frequency_analysis}-(a) (also summarised in \Cref{tab:delta_f}) shows the difference in stride frequency $\Delta f$ between the members of dyads. We observe a decrease in the mean difference in stride frequency with increasing interaction, with the mean difference ranging from $0.22$~Hz (close to the baselines) for non-interacting dyads to $0.13$~Hz for strongly interacting dyads. A Kruskal-Wallis test reveals a significant effect of the level of interaction on the difference in stride frequency between pedestrians ($p = 5.86 \times 10^{-3}$).

To further investigate the effect of the level of interaction on gait synchronisation, we perform a Dunn's test with a Bonferroni correction (see \Cref{tab:dunn_delta_f}). The test reveals a significant difference in the difference in stride frequency between dyads with intensity level 2 compared to the baselines ($p < 10^{-4}$ and $p = 3.10 \times 10^{-4}$ for $B_r$ and $B_c$, respectively). The difference in stride frequency between dyads with intensity level 3 and the baselines is also significant ($p < 10^{-4}$ and $p = 3.58 \times 10^{-4}$ for $B_r$ and $B_c$, respectively), as well as the difference between dyads with intensity level 3 and dyads with intensity level 0 ($p = 1.98 \times 10^{-2}$).

We also compute the Strictly Standardised Mean Difference (SSMD) to quantify the effect size of the difference in stride frequency between dyads and the baselines (see \Cref{tab:ssmd_delta_f}). We observe a trend of larger effect sizes with differences in the interaction level, with the SSMD ranging from $-4.89 \times 10^{-3}$ for 1--2 to $5.20 \times 10^{-1}$ for $B_c$--3.

\begin{figure}
    \centering
    \includegraphics[width=\textwidth]{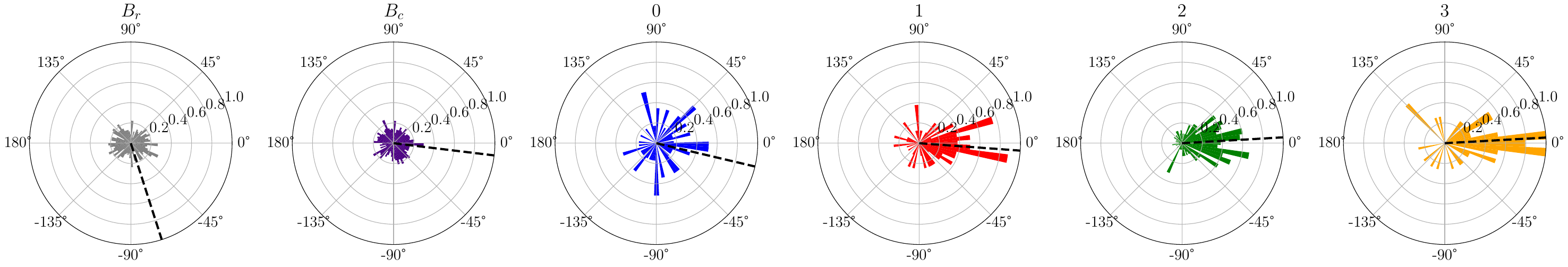}
    \caption{Mean relative phase. Polar histogram of the mean relative phase for different levels of interaction and baselines.}
    \label{fig:relative_phase_interaction}
\end{figure}

\input{tables/relative_phase.tex}

In \Cref{fig:relative_phase_interaction}, we present a polar histogram of the mean relative phase for the different levels of interaction, as well as for the baselines. We first observe that, for both baselines, the mean relative phase is almost uniformly distributed around the circle, indicating no preferred phase locking between pedestrians. This is expected for randomly paired pedestrians ($B_r$) since they are completely independent of each other and could only synchronise by chance. For the baseline $B_c$, we observe a similar distribution, indicating that pedestrians walking close to each other but not part of the same dyad may not synchronise their gait. For the different levels of interaction, we observe a clear trend of phase locking with a preferred relative phase around $0$ radians, indicating that pedestrians tend to synchronise  their gait in-phase.

The values of the circular mean of the relative phases and circular variance (as defined in \Cref{sec:relative_phase}) are summarised in \Cref{tab:relative_phase}. We observe that the mean relative phase gets closer to $0\degree$ as the level of interaction increases, with the mean relative phase ranging from $-13.30\degree$ for non-interacting dyads to $3.20\degree$ for strongly interacting dyads. The circular variance also decreases with the level of interaction, ranging from $0.75$ for non-interacting dyads to $0.35$ for strongly interacting dyads. The circular variance for the baselines are $0.98$ and $0.94$, indicating a distribution close to uniform around the circle (as observed in the polar histogram).

\begin{figure}
    \centering
    \includegraphics[width=\textwidth]{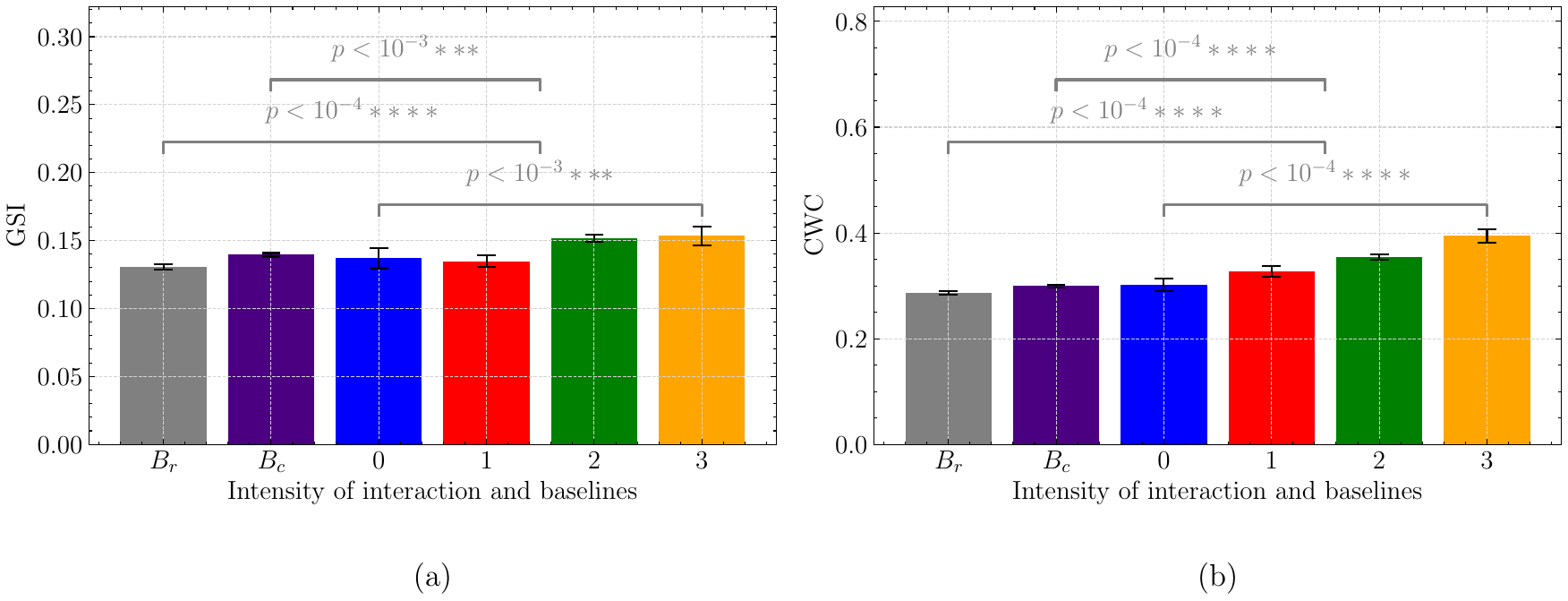}
    \caption{Effect of interaction on gait synchronisation. (a) Mean GSI values for different levels of interaction and baselines. (b) Mean coherence values for different levels of interaction and baselines.}
    \label{fig:sync_bar}
\end{figure}

In \Cref{fig:sync_bar}-a, we present the mean GSI values for different levels of interaction, as well as the baselines, and the corresponding values are summarised in \Cref{tab:sync_stats}. We observe a trend of increasing GSI with increasing level of interaction, with the mean GSI values ranging from $0.13$ for  weakly interacting dyads to $0.15$ for strongly interacting dyads.

To further investigate the effect of the level of interaction on gait synchronisation, we perform a statistical analysis using a Kruskal-Wallis test. The test reveals a significant effect of the level of interaction on the GSI values ($p = 7.98 \times 10^{-4}$).

Additionally, the baseline values are $0.13$ for $B_r$, which is lower than or equal to the GSI values across all levels of interaction, and $0.14$ for $B_c$, which is higher than the values for dyads with an interaction level of $1$, but lower than levels of interaction of 2 and 3. This differences are confirmed with Student's $t$-tests, which reveal a significant difference between the GSI values for dyads compared to both baselines ($p<10^{-4}$ and $p=1.69 \times 10^{-3}$ for $B_r$ and $B_c$, respectively).

The second metric we use to quantify gait synchronisation is the CWC between the gait residuals of pedestrians. In \Cref{fig:sync_bar}-b, we present the mean CWC values for different levels of interaction, as well as the baselines, and the corresponding values are summarised in \Cref{tab:sync_stats}. We observe a similar trend to the GSI analysis, with the mean CWC values increasing with the level of interaction. The mean CWC values range from $0.30$ for non-interacting dyads to $0.39$ for strongly interacting dyads. The statistical analysis using a Kruskal-Wallis test reveals a significant effect of the level of interaction on the CWC values ($p < 10^{-4}$).

The CWC value for the two baselines are $0.29$ for $B_r$ and $0.30$ for $B_c$. These values are lower than (or equal to) the CWC values for all dyads and these differences are confirmed with Student's $t$-tests, which reveal a significant difference between the CWC values for dyads compared to both baselines (both $p<10^{-4}$).

In \Cref{tab:dunn_gsi} and \Cref{tab:dunn_coherence} we present the results of the Dunn's test with a Bonferroni correction for the GSI and CWC values, respectively. The test reveals a significant difference in the GSI values between dyads the two baselines ($p < 10^{-4}$ for both GSI and $p = 9.99 \times 10^{-3}$ for CWC). For both the GSI and CWC values, the differences between interaction level 2 and baselines and interaction level 2 and 0 are also significant. Level 3 also shows a significant difference with the $B_r$ for both GSI and CWC values, and with $B_c$ and levels 0 and 1 for the CWC values.

We also compute the SSMD to quantify the effect size of the difference in GSI and CWC values between dyads and the baselines (see \Cref{tab:ssmd_gsi} and \Cref{tab:ssmd_coherence}). Similarly to the $\Delta f$ analysis, we observe a trend of larger effect sizes with differences in the interaction level.

\input{tables/gsi_coherence_interaction.tex}
\input{tables/dunn_gsi.tex}
\input{tables/ssmd_gsi.tex}
\input{tables/dunn_coherence.tex}
\input{tables/ssmd_coherence.tex}

\subsubsection{Effect of contact}

In this section, we investigate the effect of contact on gait synchronisation. This analysis is constrained by the very limited number of dyads annotated with a contact in the dataset (only $15$ dyads, see \Cref{tab:counts_dyads}-(b)). Nonetheless, we present the results of this analysis for completeness.

\begin{figure}
    \centering
    \includegraphics[width=\textwidth]{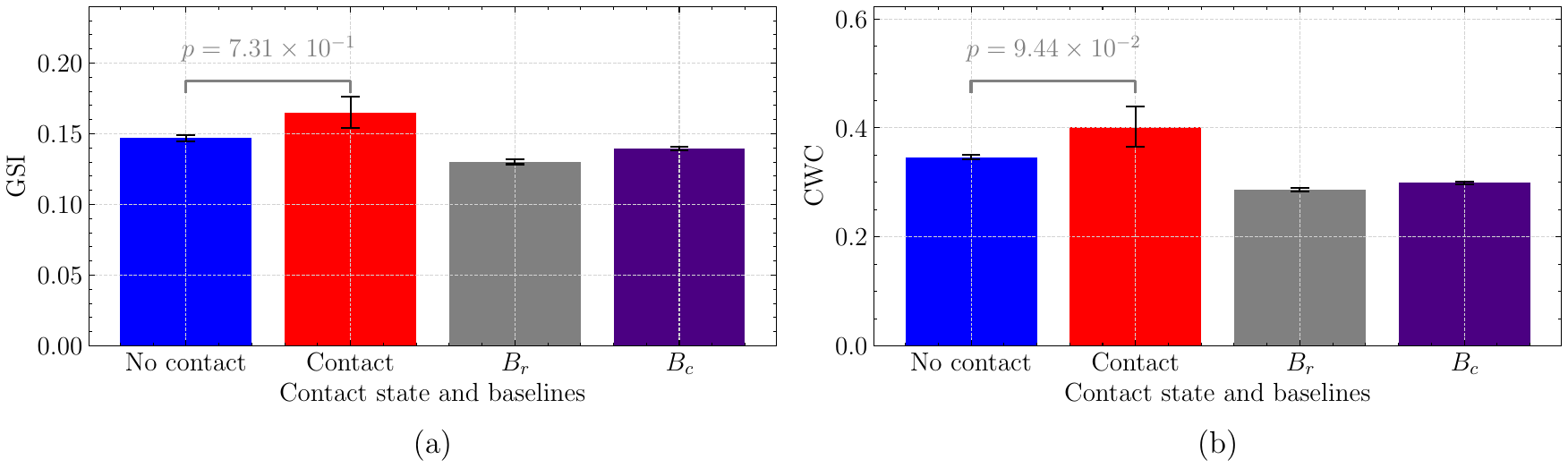}
    \caption{Effect of contact on gait synchronisation. (a) Mean GSI values for contact and no contact. (b) Mean CWC values for contact and no contact.}
    \label{fig:sync_contact}
\end{figure}

In \Cref{fig:sync_contact}, we present the mean GSI and CWC values for dyads with and without contact (also shown in \Cref{tab:sync_stats_contact}). We see that both the mean GSI and CWC values are higher for dyads with contact compared to dyads without contact ($0.15$ and $0.35$ for dyads without contact, and $0.17$ and $0.40$ for dyads with contact, respectively). Nonetheless, the statistical analysis using a Student $t$-test reveals no significant difference neither between the GSI values for dyads with and without contact ($p = 2.93 \times 10^{-1}$), nor between the CWC values ($p = 9.88 \times 10^{-2}$).

\input{tables/gsi_coherence_contact.tex}

\subsubsection{Effect of distance}

Although to our knowledge there is no direct evidence in the literature that the distance between pedestrians affects gait synchronisation, we investigate this hypothesis by analysing the GSI values for different distances between pedestrians.

\begin{figure}
    \centering
    \includegraphics[width=\textwidth]{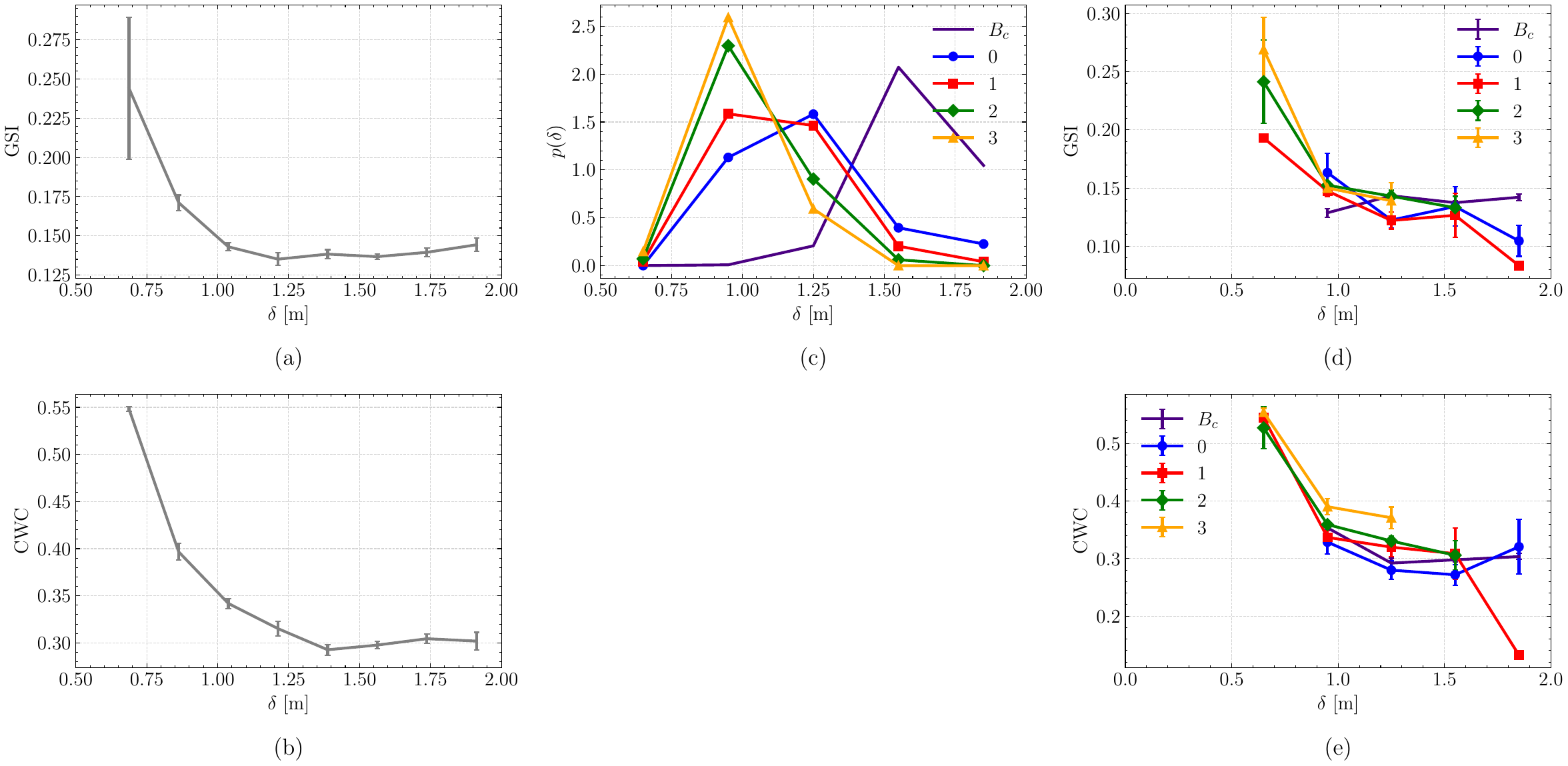}
    \caption{Gait synchronisation analysis with respect to the distance between pedestrians. (a) Binned GSI and (b) CWC values as a function of the distance $\delta$ between pedestrians in dyads and baseline $B_c$. (c) Distribution of the distance $\delta$ between dyad members with respect to the level of interaction as well as the baseline $B_c$. (d) Binned GSI and (e) CWC values as a function of the distance $\delta$ between pedestrians in dyads for different levels of interaction and baseline $B_c$.}
    \label{fig:gsi_coherence_distance}
\end{figure}

In \Cref{fig:gsi_coherence_distance}-a and \Cref{fig:gsi_coherence_distance}-b, we present the average GSI and CWC values binned with respect to the distance between pedestrians (in both dyads and baseline $B_c$). We observe a clear trend of decreasing GSI and CWC with increasing distance, indicating that pedestrians tend to synchronise their gait more when they are closer to each other.

Nonetheless, its important to note that for the dyads, the distance between pedestrians is not independent of the level of interaction. In \Cref{fig:gsi_coherence_distance}-c, we present the distribution of the distance between dyad members for different levels of interaction. We observe that higher levels of interaction are correlated with closer distances between pedestrians. This dependence has even been modelled in previous studies~\cite{zanlungo2014potential,yucel2018modeling}. Therefore, it is important to investigate the effect of these two factors independently to understand their individual contributions to gait synchronisation. In \Cref{fig:gsi_coherence_distance}-d and \Cref{fig:gsi_coherence_distance}-e, we present the average GSI and CWC values binned with respect to the distance between pedestrians in dyads as well as the baseline $B_c$, for different levels of interaction.

We observe that the correlation between distance and GSI and CWC values is still present when considering the level of interaction, with larger distance leading to lower GSI and CWC values for all levels of interaction.

It is harder to discern the effect of interaction level at specific distances due to reduced sample sizes, which result in empty bins and greater variance in the data. Despite these limitations, interaction level 3 consistently shows higher CWC values compared to other levels of interaction. Interaction level 2 also demonstrates consistently higher GSI values than interaction level 1, although for CWC, the ordering shifts in certain bins. For CWC, non-interacting dyads consistently exhibit lower values than the other interaction levels in all bins except $[1.7, 2.0]$~m.

The values for baseline $B_c$ are presented in \Cref{fig:gsi_coherence_distance}-c, d, e. Based on the probability density function of interpersonal distances (\Cref{fig:gsi_coherence_distance}-c), pairs of unrelated individuals are rarely found at distances below $1.2$~m, whereas such distances are common for dyads. Consequently, GSI and CWC values for the baseline $B_c$ could not be computed for bins below $1$~m. For bins above $1$~m, the GSI and CWC values for the baseline $B_c$ are consistently lower than those for strongly interacting dyads (interaction level 3) but higher than those for non-interacting dyads (interaction level 0). For intermediate interaction levels, the baseline $B_c$ values alternate between being higher or lower.

\subsubsection{Nonlinear analysis}

In this section, we present the results of the nonlinear analysis. We investigate the determinism, maximal Lyapunov exponent, and Cross Recurrence Analysis of pedestrian dyads and baselines.

\begin{figure}
    \centering
    \includegraphics[width=\textwidth]{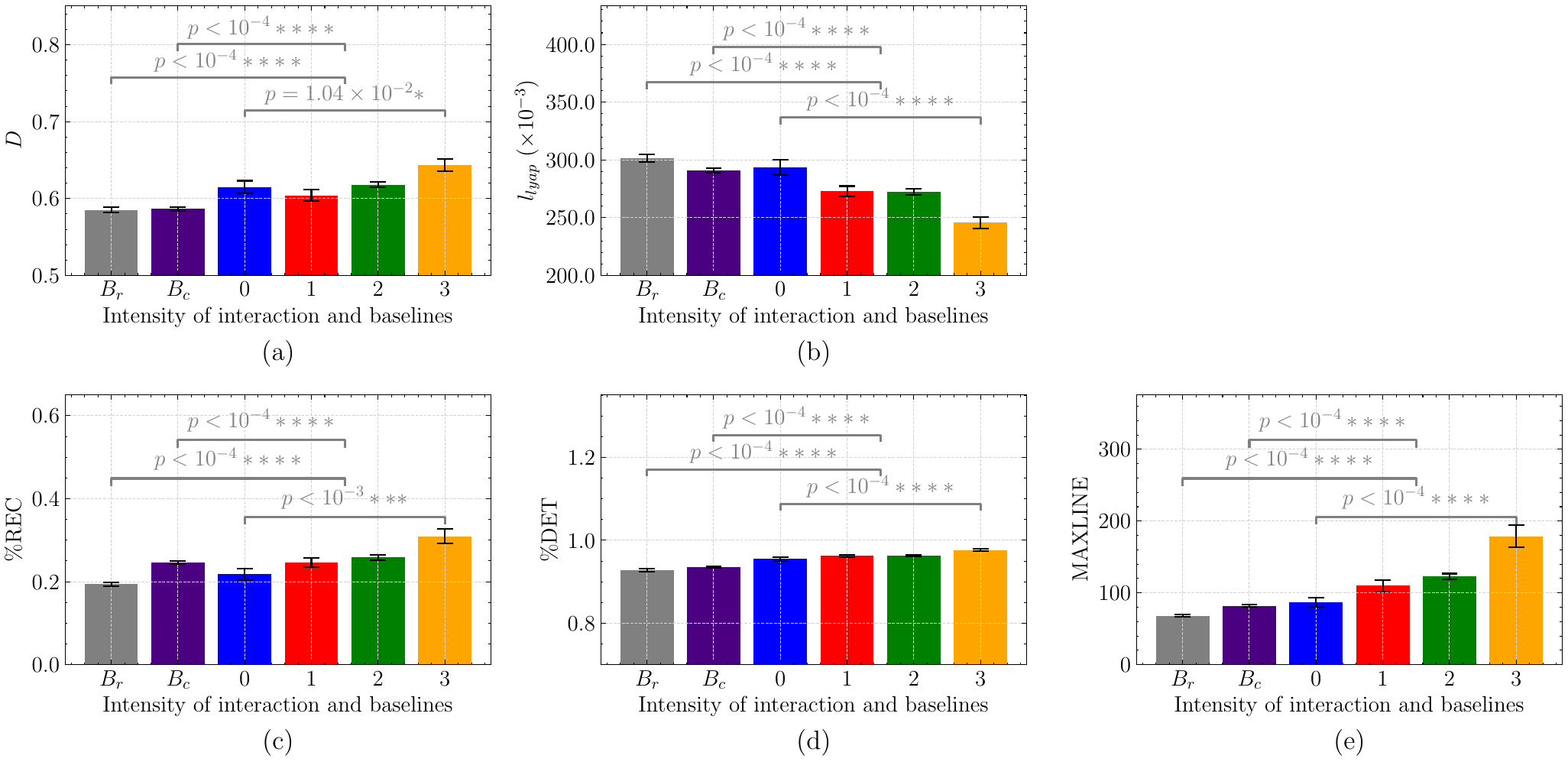}
    \caption{Nonlinear analysis of pedestrian dyads. (a) Determinism values $D$, (b) maximal  Lyapunov exponent $l_{lyap}$, (c) percentage of recurrence $\%\text{REC}$, (d) percentage of determinism $\%\text{DET}$, and (e) maximal line length $\text{MAXLINE}$ for different levels of interaction and baselines.}
    \label{fig:result_nonlinear}
\end{figure}

In \Cref{fig:result_nonlinear}-a, we present the determinism values for different levels of interaction, as well as baseline determinism values. The determinism values are all low, ranging from $0.59$ for the baselines to $0.64$ for strongly interacting dyads. There is a trend of increasing determinism with increasing interaction, with a Kruskal-Wallis test revealing significant effect of the level of interaction on the determinism values ($p = 1.04 \times 10^{-2}$).

The baseline values of determinism for random pairs of pedestrians ($B_r$) and pairs of pedestrians walking close to each other ($B_c$) are lower than the values for dyads for all levels of interaction and these differences are confirmed with Student's $t$-tests, which reveal a significant difference between the values for dyads compared to both baselines ($p<10^{-4}$ for both $B_r$ and $B_c$).

In \Cref{fig:result_nonlinear}-b, we present the maximal Lyapunov exponent values of pedestrians in dyads with different levels of interaction and two baselines maximal  Lyapunov exponent value for randomly paired pedestrians. We observe a significant ($p<10^{-4}$) effect of the level of interaction on the maximal  Lyapunov exponent values, where stronger interaction is associated with lower values of the maximal Lyapunov exponent. At all interaction levels, the maximal  Lyapunov exponent values are lower than the baseline $B_r$ and only interaction level 0 has a higher value than the baseline $B_c$. These differences are confirmed with Student's $t$-tests, which reveal a significant difference between the values for dyads compared to both baselines ($p<10^{-4}$ for both $B_r$ and $B_c$).

In \Cref{fig:result_nonlinear}-c, d, e, we present the results of CRA. We show the percentage of recurrence points $\%\text{REC}$, the percentage of determinism $\%\text{DET}$, and the maximal line length $\text{MAXLINE}$ for different levels of interaction, as well as the two baselines. We observe a significant effect of the level of interaction on all three metrics, with higher levels of interaction associated with higher values of $\%\text{REC}$, $\%\text{DET}$, and $\text{MAXLINE}$.

We also observe that the baseline values for random pairs of pedestrians ($B_r$) are lower than the values for dyads for all three metrics. The baseline value for pairs of pedestrians walking close to each other ($B_c$) is generally higher than the $B_r$ values, but lower than the values for dyads, except for $\%\text{REC}$, where the values are higher than interaction level 0. These differences are confirmed with Student's $t$-tests, which reveal a significant difference between the values for dyads compared to both baselines ($p<10^{-4}$ for both $B_r$ and $B_c$).

In \Crefrange{tab:dunn_determinism}{tab:dunn_maxline} we present the results of the Dunn's test with a Bonferroni correction for the determinism, maximal Lyapunov exponent, $\%\text{REC}$, $\%\text{DET}$, and $\text{MAXLINE}$ values, respectively. In \Crefrange{tab:ssmd_determinism}{tab:ssmd_maxline} we present the SSMD values for these metrics. Both the Dunn's test and the SSMD values confirm the results of the Kruskal-Wallis test and the Student's $t$-tests, showing a significant difference between values with large difference in the interaction level.

\input{tables/dunn_determinism.tex}
\input{tables/ssmd_determinism.tex}
\input{tables/dunn_lyapunov.tex}
\input{tables/ssmd_lyapunov.tex}
\input{tables/dunn_rec.tex}
\input{tables/ssmd_rec.tex}
\input{tables/dunn_det.tex}
\input{tables/ssmd_det.tex}
\input{tables/dunn_maxline.tex}
\input{tables/ssmd_maxline.tex}

\subsubsection{Synchronisation in triads}

In this section, we investigate gait synchronisation in triads. Similar to the dyads, we compute the GSI and CWC values between pairs of pedestrians in the triads.

We start by considering only the possible effect of the formation, i.e.\ we average the synchronisation metrics over all three pairs in the triad, and first investigate how it affects the mean difference in stride frequency $\Delta f$. We described our classification of the relative positioning of pedestrians in a triad into four categories ($\vee$, $\wedge$, $\longleftrightarrow$, and $\updownarrow$) in \Cref{sec:triads}. Since only one triad was classified as $\updownarrow$, we exclude it from the analysis.

\begin{figure}
    \centering
    \includegraphics[width=0.6\textwidth]{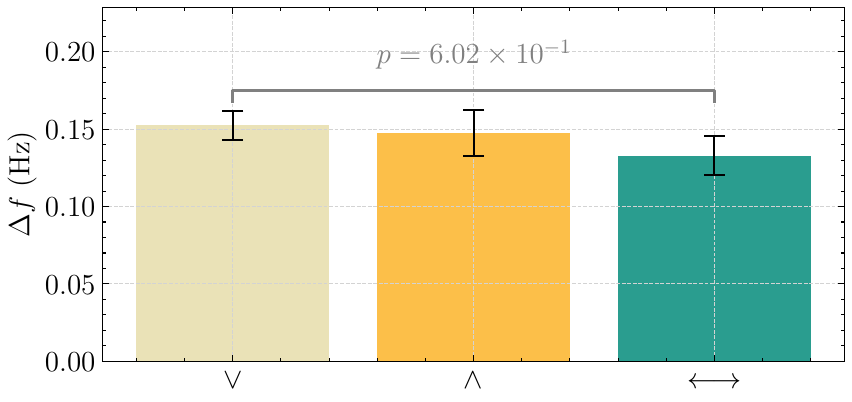}
    \caption{Mean difference in stride frequency $\Delta f$ for different formations. The error bars represent the standard error of the mean.}
    \label{fig:delta_f_formation}
\end{figure}

In \Cref{fig:delta_f_formation}, we show the mean difference in stride frequency $\Delta f$ for different formations. We observe that the $\longleftrightarrow$ formation has the smallest mean difference in stride frequency, followed by the $\wedge$ formation, and the $\vee$ formation has the largest mean difference in stride frequency. Nonetheless, a Kruskal-Wallis test reveals a nonsignificant effect of the formation on the difference in stride frequency ($p = 6.02 \times 10^{-1}$).

\begin{figure}
    \centering
    \includegraphics[width=\textwidth]{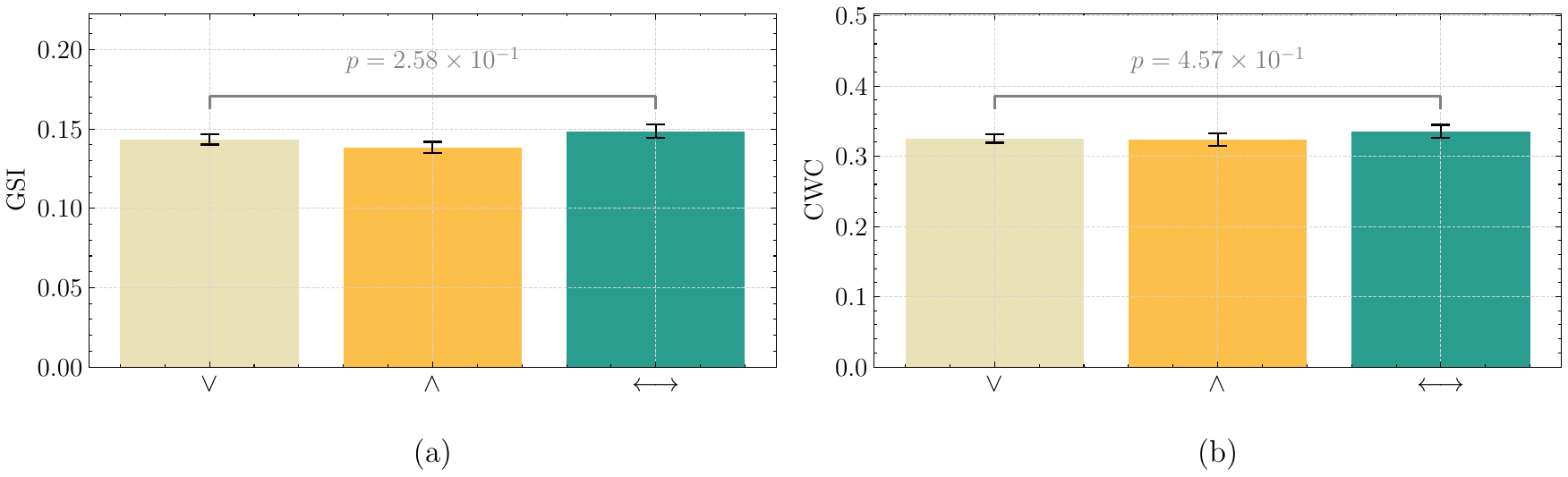}
    \caption{Gait synchronisation analysis in triads. (a) Mean GSI  and (b) CWC values for different formations. The error bars represent the standard error of the mean.}
    \label{fig:gsi_coherence_formation}
\end{figure}

In \Cref{fig:gsi_coherence_formation} we consider the average GSI and CWC values for the various formations. Consistently with $\Delta f$, we find that the $\longleftrightarrow$ formation has the highest GSI and CWC values. But again, the Kruskal-Wallis test reveals a nonsignificant effect of the formation on these metrics ($p = 2.58 \times 10^{-2}$ and $p = 4.57 \times 10^{-1}$).

\begin{figure}
    \centering
    \includegraphics[width=\textwidth]{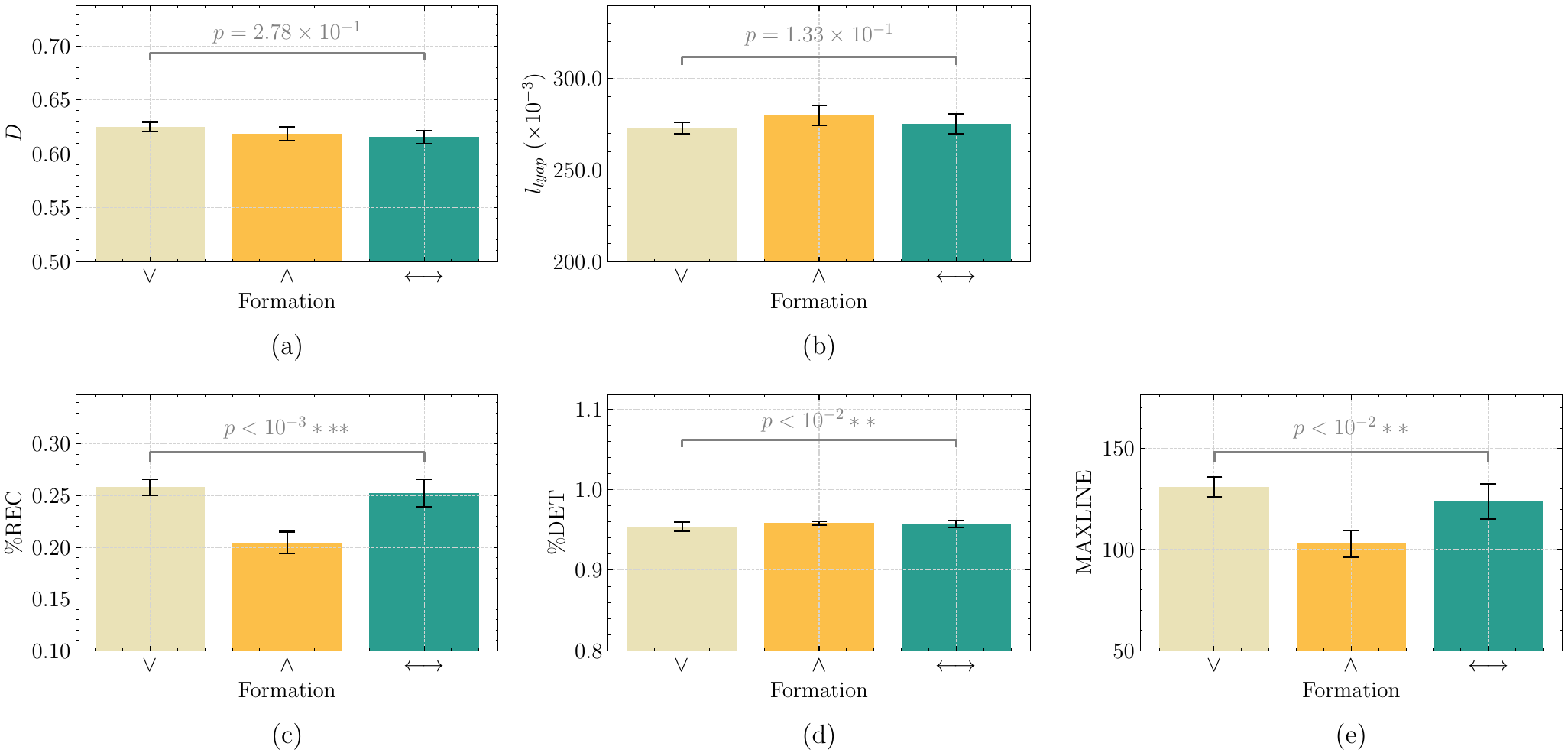}
    \caption{Nonlinear analysis of pedestrian triads. (a) Determinism values $D$, (b) maximal Lyapunov exponent values $l_{lyap}$, (c) percentage of recurrence $\%\text{REC}$, (d) percentage of determinism $\%\text{DET}$, and (e) maximal line length $\text{MAXLINE}$ for different formations. The error bars represent the standard error of the mean.}
    \label{fig:triads_nonlinear_metrics}
\end{figure}

We also perform a nonlinear analysis of the pedestrian triads. We compute the determinism $D$, maximal Lyapunov exponent $l_{lyap}$, percentage of recurrence $\%\text{REC}$, percentage of determinism $\%\text{DET}$, and maximal line length $\text{MAXLINE}$ for the different formations. The results are presented in \Cref{fig:triads_nonlinear_metrics}.

For $D$, the values are very similar for the different formations. A Kruskal-Wallis test reveals a nonsignificant effect of the formation on the determinism values ($p = 2.78 \times 10^{-1}$). For the maximal Lyapunov exponent $l_{lyap}$, we observe that the $\vee$ formation has the lowest values, followed by the $\longleftrightarrow$ formation, and the $\wedge$ formation has the highest values. This difference is nonsignificant according to the Kruskal-Wallis test ($p = 1.33 \times 10^{-1}$). For $\%\text{REC}$, $\%\text{DET}$, and $\text{MAXLINE}$, we observe that the $\wedge$ formation consistently exhibits the smallest values, while the $\longleftrightarrow$ and $\vee$ formations have higher and similar values. Kruskal-Wallis tests reveal a significant effect of the formation on these metrics ($p < 1.36 \times 10^{-4}$, $p = 3.81 \times 10^{-3}$, and $p = 1.02 \times 10^{-3}$, respectively).

\begin{figure}
    \centering
    \includegraphics[width=0.6\textwidth]{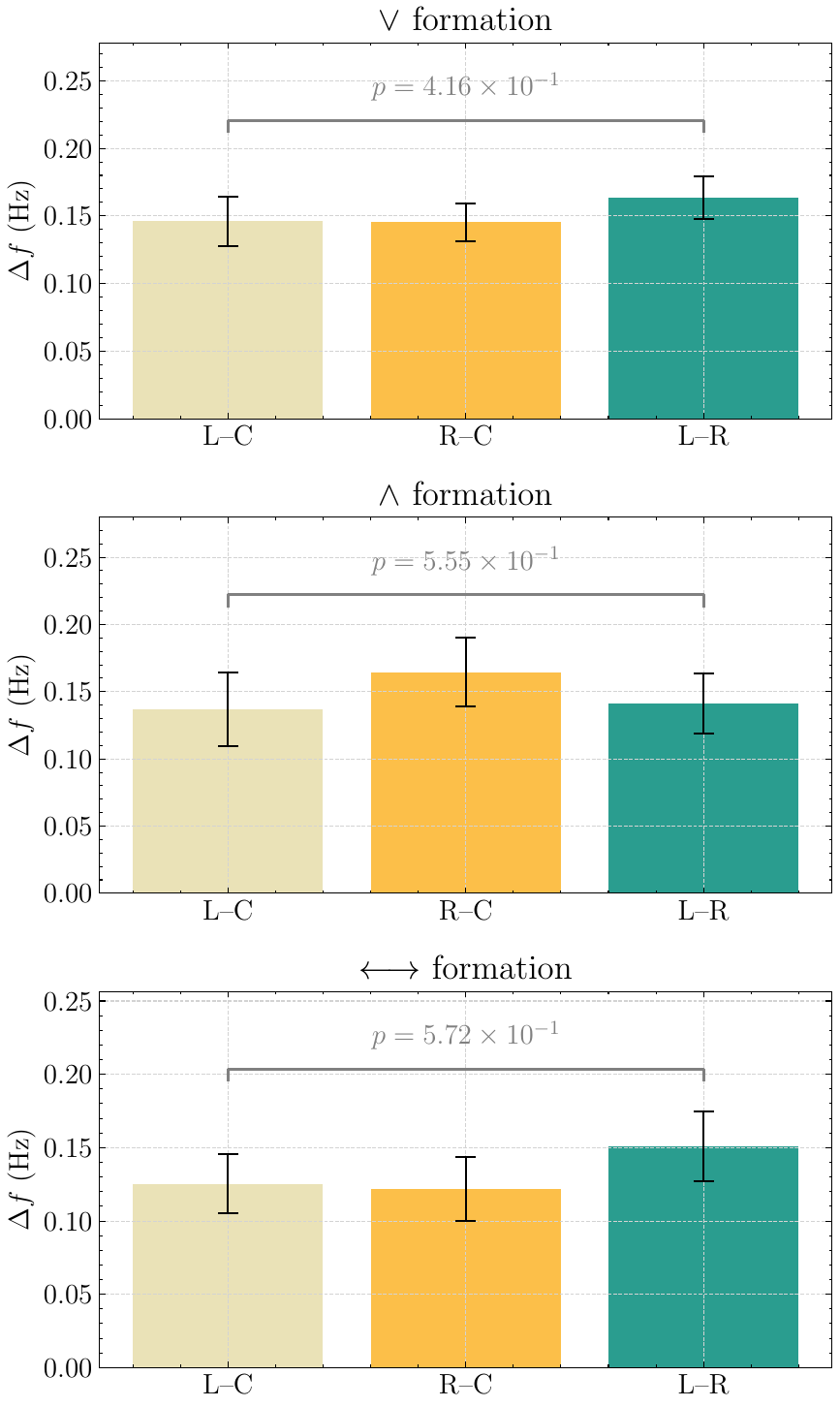}
    \caption{Mean difference in stride frequency $\Delta f$ for different relative positions in triads and formations. The error bars represent the standard error of the mean.}
    \label{fig:delta_f_position}
\end{figure}

\begin{figure}
    \centering
    \includegraphics[width=\textwidth]{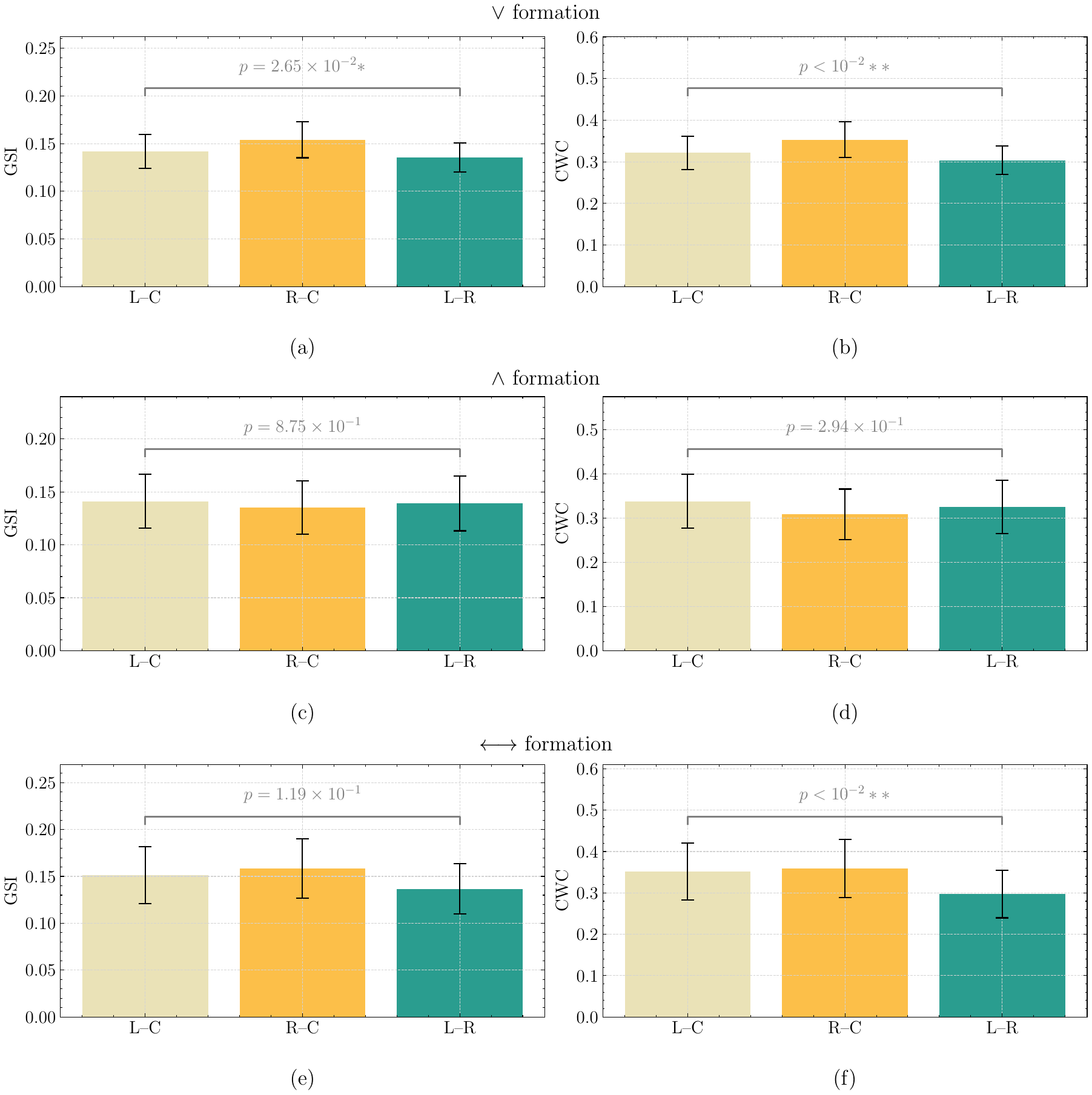}
    \caption{Gait synchronisation analysis in triads. (a) Mean GSI and (b) CWC values for different relative positions in triads in the $\vee$ formation. (c, d) Same metrics for the $\wedge$ formation. (e, f) Same metrics for the $\longleftrightarrow$ formation. The error bars represent the standard error of the mean.}
    \label{fig:gsi_coherence_position}
\end{figure}

Finally, we investigate (in \Cref{fig:delta_f_position} and \Cref{fig:gsi_coherence_position}) the effect of the relative positioning of pedestrians in the triad. For each formation, we consider the pedestrian on the left (L), the pedestrian in the centre (C), and the pedestrian on the right (R), as described in \Cref{sec:triads} and conduct the synchronisation analyses for each pair of pedestrians in the triad, L--C, C--R, and L--R (refer to \Cref{tab:counts_triads_pairs} for the number of pairs in each formation).

In \Cref{fig:delta_f_position}, we consider the mean difference in stride frequency $\Delta f$ for different relative positions in triads. It seems that there is no consistent trend in the mean difference in stride frequency for different relative positions in the triads. The L--R pair has the largest mean difference in stride frequency for the $\vee$ and $\longleftrightarrow$ formations, while the R--C pair has the largest mean difference in stride frequency for the $\wedge$ formation. Kruskal-Wallis also reveals a nonsignificant effect of the relative position on the difference in stride frequency ($p = 4.16 \times 10^{-1}$, $p = 5.55 \times 10^{-1}$, and $p = 5.72 \times 10^{-1}$ for the $\vee$, $\wedge$, and $\longleftrightarrow$ formations, respectively).

In \Cref{fig:gsi_coherence_position}, we present the GSI and CWC values for the same decomposition of the triads. For the $\vee$ formation, we observe that the R--C pair has the highest GSI and CWC values, while the L--R pair has the lowest values, and that Kruskal-Wallis found these differences to be significant ($p=2.65 \times 10^{-2}$ and $p=2.30 \times 10^{-3}$ respectively). For the $\wedge$ formation, the values are closer to each other, and no significant differences were found. The $\longleftrightarrow$ formation shows a similar trend to the $\vee$ formation, with the R--C pair having the highest GSI and CWC values, and the L--R pair having the lowest values. Kruskal-Wallis found these differences to be significant only for the CWC values ($p=7.40 \times 10^{-3}$).

\section{Discussion}

Our study provides new insights into the influence of social interaction on pedestrian gait dynamics and synchronisation. We have used a large ecological dataset of pedestrian trajectories to investigate gait synchronisation between pedestrians in dyads and triads, as well as the effect of the level of interaction, contact, and distance between pedestrians on gait coordination.

Our results on gait synchronisation demonstrate the importance of interaction level in modulating dyadic synchronisation. Higher interaction levels were associated with reduced differences in stride frequency (see \Cref{fig:frequency_analysis}-b) and greater in-phase synchronisation, as shown by the smaller variation in the relative phase and higher GSI and CWC values (see \Cref{fig:sync_bar} and \Cref{fig:relative_phase_interaction}).

Zivotofsky et al.~\cite{zivotofsky2018effects} reported that dual tasking can affect the synchronisation of gait patterns, with a simple dual task increasing synchronisation and a complex dual task reducing synchronisation. We argue that the interaction between pedestrians may be considered as a form of dual tasking, where the cognitive load of coordinating movements and maintaining social interaction may affect the synchronisation of gait patterns. In Zivotofsky et al.'s study, the simple dual task consisted of listening to a section from a story through headphones and paying attention to two phonemes, while the complex dual task required listening for four phonemes and the content of the story. We hypothesise that even strongly interacting dyads (level $3$) may not need to perform mental tasks as complex as these, as usual social interactions may only require to engage in a conversation without the need to pay attention to such specific details. We may then consider that levels of interaction $1$ to $3$ all correspond to simple dual tasks.

Since levels of interaction have been shown to be correlated with the distance between pedestrians, with higher interaction levels associated with closer distances (see \Cref{fig:gsi_coherence_distance}-c), we also investigated the effect of distance on gait synchronisation (see \Cref{fig:gsi_coherence_distance}-d,e). We found that closer distances were associated with higher GSI and CWC values, indicating that pedestrians tend to synchronise their gait more when they are closer to each other.

Trying to disentangle the effect of distance and interaction levels, we observed that the correlation between level of interaction and gait synchronisation metrics is less evident when considering the distance between pedestrians, but still visible for CWC values. The separation of these two factors is challenging, as it tends to lower the sample size, which may affect the reliability of the results. Trying to compare these results to a baseline of individuals walking close to each other also proved to be challenging, since situations where pedestrians are walking close to each other but not part of the same dyad are rare in a dataset with a low density of pedestrians. We found that for the baseline, the GSI and CWC also tend to decrease with increasing distance. Situations where the distance is small may correspond to cases where pedestrians are overcoming one another. While they overcome, individuals may need to synchronise their gait with the pedestrian they are overtaking. Since they are not cognitively engaged in interaction, they may be able to synchronise successfully, leading to higher GSI and CWC values.

Interestingly, although we did find higher values of GSI and CWC for dyads with contact compared to dyads without contact, the difference was not significant. Other studies~\cite{harrison2009horsing,sylos-labini2018human} have reported statistically significant differences in gait synchronisation between dyads with and without contact. We argue that the main reason behind the  lack of significance in our study may be the limited number of dyads with contact in our dataset (only $15$ dyads annotated with contact). In addition, the definition of contact in our study is not restrained to hand-holding (or other coupling involving maintained physical contact), but may also include other forms of temporally limited contact, such as brushing shoulders.

The nonlinear analysis revealed small but significant differences in the determinism of the reconstructed phase space of the gait residuals between dyads with different levels of interaction (see \Cref{fig:result_nonlinear}-a). The determinism values were higher for dyads with stronger interactions, suggesting that their gait patterns exhibit more predictable and structured dynamics.

Regarding the maximal Lyapunov exponent values, they tended to decrease with stronger interactions, suggesting less chaotic gait patterns for dyads with stronger interactions (see \Cref{fig:result_nonlinear}-b). A lower Lyapunov exponent indicates greater stability in the system's dynamics, meaning that small perturbations in gait trajectories do not amplify as rapidly. This finding aligns with the idea that social interaction may lead to more stable and synchronised movement patterns. Together, these results support the notion that social interaction can shape the dynamics of pedestrian gait patterns.

Cross Recurrence Analysis further corroborated the role of interaction levels in shaping dyadic gait patterns. The recurrence rate, determinism, and maximal line length were significantly higher in strongly interacting dyads compared to weaker interactions or baseline conditions (see \Cref{fig:result_nonlinear}-c,d,e).

We also extended our analysis to triads, investigating the effect of the formation (see \Cref{fig:delta_f_formation} and \Cref{fig:gsi_coherence_formation}) or relative positioning (see \Cref{fig:delta_f_position} and \Cref{fig:gsi_coherence_position}) on gait synchronisation. In addition to the small sample size of triads in our dataset that may have limited the statistical power of our analysis, the complexity of the interactions in triads may have introduced additional variability that made it difficult to identify clear patterns. We could not find a consistent effect of the formation or relative positioning on gait synchronisation in triads using the GSI and CWC metrics. Nonetheless, the nonlinear analysis revealed significant differences in the maximal Lyapunov  exponent, percentage of recurrence, percentage of determinism, and maximal line length between different formations, suggesting that the triad in a $\wedge$ exhibits less structured and more chaotic gait patterns compared to the $\vee$ and $\longleftrightarrow$ formations (see \Cref{fig:triads_nonlinear_metrics}). Looking at the distribution of relative position in the triads (see \Cref{fig:heatmap_positions}), we observe that for the $\wedge$ formation, the positions are more spread out than for the $\vee$ and $\longleftrightarrow$ formations. This means that the $\wedge$ formation may be less stable than the other formations, with pedestrians being more likely to slow down or speed up to maintain their relative positions in the triad, possibly switching to different formations as they walk. Zanlungo et al.~\cite{zanlungo2015spatial-size} have shown that the $\vee$ formation is very stable, regardless of the density of pedestrians, while the $\wedge$ formation is more frequent at higher densities. In the later, pedestrians may switch to collision avoidance strategies, which may introduce additional variability in the gait patterns, leading to the observed differences in the nonlinear metrics.

Looking at pairs of pedestrians in the triads, we found that there is a tendency for the L--R pair to have lower GSI and CWC values compared to the R--C and L--C pairs in the $\vee$ and $\longleftrightarrow$ formations (see \Cref{fig:gsi_coherence_position}). This might be explained by the fact that the L--R pair is the most distant pair in the triad and that these two pedestrians may actually not have a direct interaction (especially in the $\longleftrightarrow$, since the centre pedestrian is is an immediate neighbour to both of the remaining two members of the triad).

In this study we have investigated multiple aspects of gait synchronisation in non-instructed pedestrian groups, leveraging ecological data and a diverse set of analytical tools. We applied various methodologies, including the Gait Synchronisation Index, Cross-Wavelet Coherence, and several nonlinear analysis techniques.

While all methods fundamentally converged on the same overall findings, certain analyses---particularly those using CWC, Lyapunov exponent, $\text{MAXLINE}$, and $\Delta f$---were more effective in highlighting differences between conditions. These methods often yielded larger effect sizes and statistically significant results, even in cases where other approaches either failed to detect the same differences or did so without reaching statistical significance.

Our findings suggest that these metrics may be particularly well-suited for studying gait synchronisation in ecological settings, especially when employing our proposed oversmoothing method to extract gait residuals.

\section{Conclusion}

This study contributes to the literature both methodologically, by introducing a novel detection method for gait synchronisation, and empirically, by offering insights into the emergence and  stability of spontaneous gait synchronisation as well as the social factors that influence it.

Our most important methodological contribution focuses on deriving gait oscillations from trajectory data. Specifically, we propose a technique based on oversmoothing of pedestrian trajectories, allowing us to compute the difference between the original trajectory and its oversmoothed counterpart, which we refer to as gait residuals. These residuals capture the oscillations generated by the body's movement between left and right stances, facilitating time and frequency analyses that can provide insights into the temporal evolution and phase-locking patterns of gait. Our gait frequency analyses have demonstrated that the proposed methodology is effective in capturing the gait patterns of pedestrians, since the gait parameters extracted from the dataset align with established norms in the literature.

Another methodological contribution relates the deployment of  cross recurrence analysis in assessing stability of gait synchronisation and the effect of various social factors on it.
While cross recurrence analysis was introduced some time ago, it has, to the best of our knowledge, not yet been applied to gait data. In this study, we demonstrate that gait data is particularly well-suited for this type of analysis, revealing valuable insights in the process.

Our empirical contributions arise from the investigation of the impact of social interaction on pedestrian gait synchronisation using a large ecological dataset of pedestrian trajectories. We  analyse  gait dynamics of pedestrians in dyads and triads, focusing on the effect of the level of interaction, physical contact, and distance between pedestrians on gait coordination.
As a result of this analysis, we provide robust evidence of the impact of social interaction on pedestrian gait dynamics. Higher interaction levels lead to greater gait synchronisation and increased predictability, highlighting the interplay between social coordination and gait patterns. These findings add to a growing body of literature on social locomotion~\cite{gregorj2024social2,gregorj2023social1,gregorj2024asymmetries,gregorj2024ecological} .

Our findings offer significant potential for applications across a range of fields, including medical diagnosis, assistive technologies, robotics, and urban planning. In particular, we see two promising applications in medical diagnosis and rehabilitation therapy that are closely aligned with our methodological contributions. For instance, analysing an individual's gait residuals can reveal the degree of symmetry in their locomotion, helping to determine whether they exhibit equal lateral movement or asymmetrical gait patterns. Such assessment can serve as an early indicator of certain neurodegenerative diseases. Additionally, for individuals undergoing rehabilitation after health events such as strokes, evaluating gait symmetry alongside other metrics can provide a quantitative measure of therapy effectiveness. Furthermore, the concepts of gait synchronisation and symmetry may be integrated, as existing literature suggests that paired walking exercises can enhance rehabilitation outcomes. A deeper understanding of the mechanisms behind human gait synchronisation could also inform the development of assistive technologies for individuals with gait impairments, such as exoskeletons or prosthetic devices. While bipedal robots that share public spaces with humans are still in their early stages, our findings may guide the design of such robots so that they can adapt their gait patterns to synchronize with their human counterparts.

As often the case with observational studies, our analysis is limited by the constraints of the dataset. The dataset used in this study was collected in an ecological setting, which may introduce confounding factors that are difficult to control (e.g.\ varying pedestrian densities, environmental conditions, or cultural norms). In addition, the annotations of the level of interaction and physical contact are subjective (although they were performed by multiple annotators to ensure reliability) and may contain errors or biases.

In the future, the results of this study could be validated in controlled experiments. Virtual Reality (VR) simulations have been used to study paired walking in a controlled environment~\cite{soczawa2020gait}, and could also be extended to investigate the effect of social interaction on gait synchronisation. Future work could also investigate the impact of other factors on gait synchronisation, such as the social relation between pedestrians (e.g.\ couples, colleagues, etc.).

%The ATC dataset contains such information, but unfortunately, the time resolution of the trajectories is not sufficient to capture the gait dynamics of pedestrians. We hope that future datasets will provide more detailed information on the social relations between pedestrians, as well as the gait dynamics of pedestrians.

\printbibliography

\end{document}

%% file: tables/counts_dyads.tex
\begin{table}
\centering
\caption{Breakdown of the number of dyads for (a) different intensities of interaction and (b) presence of contact.}
\label{tab:counts_dyads}
\begin{subtable}{.45\linewidth}
\centering
\caption{}
\begin{tabular}{lc}
\toprule
Intensity of interaction & Count \\
\midrule
Interaction 0 & 63 \\
Interaction 1 & 94 \\
Interaction 2 & 377 \\
Interaction 3 & 75 \\
\bottomrule
\end{tabular}
\end{subtable}
\begin{subtable}{.45\linewidth}
\centering
\caption{}
\begin{tabular}{lc}
\toprule
Contact & Count \\
\midrule
No contact & 594 \\
Contact & 15 \\
\bottomrule
\end{tabular}
\end{subtable}
\end{table}

%% file: tables/counts_triads.tex
\begin{table}
\centering
\caption{Breakdown of the number of triads for each formation.}
\label{tab:counts_triads_formations}
\begin{tabular}{lcccccc}
\toprule
Formation & Count \\
\midrule
$\vee$ & 91 \\
$\wedge$ & 34 \\
$\longleftrightarrow$ & 31 \\
$\updownarrow$ & 1 \\
\bottomrule
\end{tabular}
\end{table}

%% file: tables/counts_triads_pairs.tex
\begin{table}
\centering
\caption{Breakdown of the number of triads for each possible pair in the different formations. L is left, R is right, C is center, F is front, and B is back.}
\label{tab:counts_triads_pairs}
\begin{tabular}{lcccccc}
\toprule
Formation & L--C & R--C & L--R & F--C & F--B & B--C \\
\midrule
$\vee$ & 91 & 91 & 91 & - & - & - \\
$\wedge$ & 34 & 34 & 34 & - & - & - \\
$\longleftrightarrow$ & 31 & 31 & 31 & - & - & - \\
$\updownarrow$ & - & - & - & 1 & 1 & 1 \\
\bottomrule
\end{tabular}
\end{table}

%% file: tables/gait_stats.tex
\begin{table}

    \centering
    \caption{Mean value and standard deviation of velocity $v$, stride frequency $f$, and stride length $l$ for different intensities of interaction. Kruskal-Wallis $p$-values for the difference between the intensities of interaction and Student's $t$-test $p$-values for the difference between all dyads and individuals are also shown.}
    \label{tab:gait_stats}
    \begin{tabular}{lcccccc}
        \toprule
        Intensity of interaction     & \multicolumn{2}{c}{$v$ [m/s]} & \multicolumn{2}{c}{$f$ [Hz]}          & \multicolumn{2}{c}{$l$ [m]}                                                                                                         \\
        \midrule
        Interaction 0                & $1.30 \pm 0.17$               & \multirow{4}{*}{$\mathbf{< 10^{-4}}$} & $1.09 \pm 0.20$             & \multirow{4}{*}{$1.06 \times 10^{-1}$} & $1.51 \pm 0.69$      & \multirow{4}{*}{$\mathbf{< 10^{-4}}$} \\
        Interaction 1                & $1.23 \pm 0.16$               &                                       & $1.05 \pm 0.16$             &                                        & $1.39 \pm 0.44$      &                                       \\
        Interaction 2                & $1.17 \pm 0.17$               &                                       & $1.05 \pm 0.18$             &                                        & $1.43 \pm 0.51$      &                                       \\
        Interaction 3                & $1.04 \pm 0.25$               &                                       & $1.04 \pm 0.15$             &                                        & $1.20 \pm 0.34$      &                                       \\
        \midrule
        All                          & $1.18 \pm 0.20$               &                                       & $1.05 \pm 0.18$             &                                        & $1.41 \pm 0.52$      &                                       \\
        Individuals                  & $1.29 \pm 0.21$               &                                       & $1.03 \pm 0.19$             &                                        & $1.86 \pm 1.09$      &                                       \\
        \midrule
        Student's $t$-test $p$-value & $\mathbf{< 10^{-4}}$          &                                       & $1.80 \times 10^{-1}$       &                                        & $\mathbf{< 10^{-4}}$ &                                       \\
        \bottomrule
    \end{tabular}
\end{table}

%% file: tables/pearson_correlation.tex
\begin{table}
\centering
\caption{Pearson correlation coefficient $r_{vf}$ between velocity $v$ and stride frequency $f$ and $r_{vl}$ between velocity $v$ and stride length $l$ for different intensities of interaction.}
\label{tab:pearson_correlation}
\begin{tabular}{lcc}
\toprule
Intensity of interaction & $r_{vf}$ & $r_{vl}$ \\
\midrule
Interaction 0 & $0.17$ & $0.11$ \\
Interaction 1 & $0.12$ & $0.30$ \\
Interaction 2 & $0.18$ & $0.22$ \\
Interaction 3 & $0.27$ & $0.58$ \\
\midrule
All & $0.19$ & $0.25$ \\
Individuals & $0.03$ & $0.24$ \\
\bottomrule
\end{tabular}
\end{table}

%% file: tables/delta_f.tex
\begin{table}
\centering
\caption{Mean and standard error of the difference in stride frequency $\Delta f$ between baseline pairs of $B_r$ and $B_c$ as well as dyad members for different intensities of interaction. Kruskal-Wallis $p$-value for the difference between the intensities of interaction and Student's $t$-test $p$-value for the difference between all dyads and the baseline are also shown.}
\label{tab:delta_f}
\begin{tabular}{lcc}
\toprule
Intensity of interaction & \multicolumn{2}{c}{$\Delta f$ [Hz]}  \\
\midrule
Interaction 0 &$0.22 \pm 0.02$  & \multirow{4}{*}{$\mathbf{5.86 \times 10^{-3}}$} \\
Interaction 1 &$0.18 \pm 0.02$ & \\
Interaction 2 &$0.18 \pm 0.01$ & \\
Interaction 3 &$0.13 \pm 0.02$ & \\
\midrule
$B_r$ &$0.26 \pm 0.01$ & \\
$B_c$ &$0.24 \pm 0.01$ & \\
\midrule
Student's $t$-test $p$-value for $B_r$ & $\mathbf{< 10^{-4}}$ \\Student's $t$-test $p$-value for $B_c$ & $\mathbf{< 10^{-4}}$ \\\bottomrule
\end{tabular}
\end{table}

%% file: tables/dunn_delta_f.tex
\begin{table}
\centering
\caption{Dunn post-hoc test for pairwise comparisons of the difference in stride frequency $\Delta f$ between baseline pairs of $B_r$ and $B_c$ as well as different intensities of interaction. The $p$-values are adjusted using the Bonferroni correction.}
\label{tab:dunn_delta_f}
\begin{tabular}{lcccccc}
\toprule
 & $B_r$ & $B_c$ & 0 & 1 & 2 & 3 \\
\midrule
$B_r$ & - & $4.84 \times 10^{-1}$ & $1.00$ & $9.92 \times 10^{-2}$ & $\mathbf{< 10^{-4}}$ & $\mathbf{< 10^{-4}}$ \\
$B_c$ & - & - & $1.00$ & $3.58 \times 10^{-1}$ & $\mathbf{2.68 \times 10^{-4}}$ & $\mathbf{2.87 \times 10^{-4}}$ \\
0 & - & - & - & $8.35 \times 10^{-1}$ & $4.04 \times 10^{-1}$ & $\mathbf{1.45 \times 10^{-2}}$ \\
1 & - & - & - & - & $1.00$ & $1.86 \times 10^{-1}$ \\
2 & - & - & - & - & - & $1.61 \times 10^{-1}$ \\
3 & - & - & - & - & - & - \\
\bottomrule
\end{tabular}
\end{table}

%% file: tables/ssmd_delta_f.tex
\begin{table}
\centering
\caption{SSMD for pairwise comparisons of the difference in stride frequency $\Delta f$ between different intensities of interaction.}
\label{tab:ssmd_delta_f}
\begin{tabular}{lcccccc}
\toprule
 & $B_r$ & $B_c$ & 0 & 1 & 2 & 3 \\
\midrule
$B_r$ & - & $4.86 \times 10^{-2}$ & $1.42 \times 10^{-1}$ & $2.85 \times 10^{-1}$ & $2.67 \times 10^{-1}$ & $5.20 \times 10^{-1}$ \\
$B_c$ & - & - & $8.70 \times 10^{-2}$ & $2.30 \times 10^{-1}$ & $2.14 \times 10^{-1}$ & $4.64 \times 10^{-1}$ \\
0 & - & - & - & $1.70 \times 10^{-1}$ & $1.53 \times 10^{-1}$ & $4.61 \times 10^{-1}$ \\
1 & - & - & - & - & $-4.89 \times 10^{-3}$ & $2.98 \times 10^{-1}$ \\
2 & - & - & - & - & - & $2.77 \times 10^{-1}$ \\
3 & - & - & - & - & - & - \\
\bottomrule
\end{tabular}
\end{table}

%% file: tables/relative_phase.tex
\begin{table}
\centering
\caption{Circular mean and variance of the relative phase between pedestrians for different intensities of interaction and baseline.}
\label{tab:relative_phase}
\begin{tabular}{lcc}
\toprule
Intensity of interaction & Mean relative phase (°) & Variance \\
\midrule
Interaction 0 & $-13.30$ & $0.75$ \\
Interaction 1 & $-4.20$ & $0.53$ \\
Interaction 2 & $3.31$ & $0.47$ \\
Interaction 3 & $3.20$ & $0.35$ \\
\midrule
$B_r$ & $-72.21$ & $0.98$ \\
$B_c$ & $-6.92$ & $0.94$ \\
\bottomrule
\end{tabular}
\end{table}

%% file: tables/gsi_coherence_interaction.tex
\begin{table}
\centering
\caption{GSI and CWC for different intensities of interaction. Kruskal-Wallis $p$-values for the difference between the intensities of interaction and Student's $t$-test $p$-values for the difference between all dyads and the baseline are also shown.}
\label{tab:sync_stats}
\begin{tabular}{lcccc}
\toprule
Intensity of interaction & \multicolumn{2}{c}{GSI} & \multicolumn{2}{c}{CWC} \\
\midrule
Interaction 0 & $0.14 \pm 0.06$ & \multirow{4}{*}{$\mathbf{7.98 \times 10^{-4}}$} &$0.30 \pm 0.09$ & \multirow{4}{*}{$\mathbf{< 10^{-4}}$} \\
Interaction 1 & $0.13 \pm 0.04$ & &$0.33 \pm 0.09$ & \\
Interaction 2 & $0.15 \pm 0.05$ & &$0.35 \pm 0.09$ & \\
Interaction 3 & $0.15 \pm 0.04$ & &$0.39 \pm 0.08$ & \\
\midrule
$B_r$ & $0.13 \pm 0.05$ & &$0.29 \pm 0.10$ & \\
$B_c$ & $0.14 \pm 0.05$ & &$0.30 \pm 0.09$ & \\
\midrule
All & $0.15 \pm 0.05$ & & $0.35 \pm 0.09$ & \\
\midrule
Student's $t$-test $p$-value for $B_r$ & $\mathbf{< 10^{-4}}$ & & $\mathbf{< 10^{-4}}$  &\\
Student's $t$-test $p$-value for $B_c$ & $\mathbf{1.69 \times 10^{-3}}$ &  &$\mathbf{< 10^{-4}}$ &\\
\bottomrule
\end{tabular}
\end{table}

%% file: tables/dunn_gsi.tex
\begin{table}
\centering
\caption{Dunn post-hoc test for pairwise comparisons of the GSI between baseline pairs of $B_r$ and $B_c$ as well as different intensities of interaction. The $p$-values are adjusted using the Bonferroni correction.}
\label{tab:dunn_gsi}
\begin{tabular}{lcccccc}
\toprule
 & $B_r$ & $B_c$ & 0 & 1 & 2 & 3 \\
\midrule
$B_r$ & - & $\mathbf{< 10^{-4}}$ & $1.00$ & $7.29 \times 10^{-1}$ & $\mathbf{< 10^{-4}}$ & $\mathbf{1.02 \times 10^{-3}}$ \\
$B_c$ & - & - & $1.00$ & $1.00$ & $\mathbf{1.57 \times 10^{-4}}$ & $1.63 \times 10^{-1}$ \\
0 & - & - & - & $1.00$ & $\mathbf{2.14 \times 10^{-2}}$ & $8.19 \times 10^{-2}$ \\
1 & - & - & - & - & $\mathbf{4.92 \times 10^{-2}}$ & $1.63 \times 10^{-1}$ \\
2 & - & - & - & - & - & $1.00$ \\
3 & - & - & - & - & - & - \\
\bottomrule
\end{tabular}
\end{table}

%% file: tables/ssmd_gsi.tex
\begin{table}
\centering
\caption{SSMD for pairwise comparisons of the GSI between different intensities of interaction.}
\label{tab:ssmd_gsi}
\begin{tabular}{lcccccc}
\toprule
 & $B_r$ & $B_c$ & 0 & 1 & 2 & 3 \\
\midrule
$B_r$ & - & $-1.33 \times 10^{-1}$ & $-8.63 \times 10^{-2}$ & $-6.91 \times 10^{-2}$ & $-2.99 \times 10^{-1}$ & $-3.39 \times 10^{-1}$ \\
$B_c$ & - & - & $3.56 \times 10^{-2}$ & $7.67 \times 10^{-2}$ & $-1.79 \times 10^{-1}$ & $-2.17 \times 10^{-1}$ \\
0 & - & - & - & $3.01 \times 10^{-2}$ & $-1.95 \times 10^{-1}$ & $-2.29 \times 10^{-1}$ \\
1 & - & - & - & - & $-2.64 \times 10^{-1}$ & $-3.09 \times 10^{-1}$ \\
2 & - & - & - & - & - & $-2.92 \times 10^{-2}$ \\
3 & - & - & - & - & - & - \\
\bottomrule
\end{tabular}
\end{table}

%% file: tables/dunn_coherence.tex
\begin{table}
\centering
\caption{Dunn post-hoc test for pairwise comparisons of the CWC between baseline pairs of $B_r$ and $B_c$ as well as different intensities of interaction. The $p$-values are adjusted using the Bonferroni correction.}
\label{tab:dunn_coherence}
\begin{tabular}{lcccccc}
\toprule
 & $B_r$ & $B_c$ & 0 & 1 & 2 & 3 \\
\midrule
$B_r$ & - & $\mathbf{4.66 \times 10^{-3}}$ & $3.91 \times 10^{-1}$ & $\mathbf{9.56 \times 10^{-4}}$ & $\mathbf{< 10^{-4}}$ & $\mathbf{< 10^{-4}}$ \\
$B_c$ & - & - & $7.94 \times 10^{-1}$ & $6.24 \times 10^{-2}$ & $\mathbf{< 10^{-4}}$ & $\mathbf{< 10^{-4}}$ \\
0 & - & - & - & $3.91 \times 10^{-1}$ & $\mathbf{7.23 \times 10^{-4}}$ & $\mathbf{< 10^{-4}}$ \\
1 & - & - & - & - & $7.16 \times 10^{-2}$ & $\mathbf{1.60 \times 10^{-3}}$ \\
2 & - & - & - & - & - & $7.16 \times 10^{-2}$ \\
3 & - & - & - & - & - & - \\
\bottomrule
\end{tabular}
\end{table}

%% file: tables/ssmd_coherence.tex
\begin{table}
\centering
\caption{SSMD for pairwise comparisons of the CWC between different intensities of interaction.}
\label{tab:ssmd_coherence}
\begin{tabular}{lcccccc}
\toprule
 & $B_r$ & $B_c$ & 0 & 1 & 2 & 3 \\
\midrule
$B_r$ & - & $-9.06 \times 10^{-2}$ & $-1.05 \times 10^{-1}$ & $-2.92 \times 10^{-1}$ & $-4.84 \times 10^{-1}$ & $-8.03 \times 10^{-1}$ \\
$B_c$ & - & - & $-1.43 \times 10^{-2}$ & $-2.13 \times 10^{-1}$ & $-4.17 \times 10^{-1}$ & $-7.55 \times 10^{-1}$ \\
0 & - & - & - & $-2.01 \times 10^{-1}$ & $-4.06 \times 10^{-1}$ & $-7.47 \times 10^{-1}$ \\
1 & - & - & - & - & $-2.08 \times 10^{-1}$ & $-5.41 \times 10^{-1}$ \\
2 & - & - & - & - & - & $-3.20 \times 10^{-1}$ \\
3 & - & - & - & - & - & - \\
\bottomrule
\end{tabular}
\end{table}

%% file: tables/gsi_coherence_contact.tex
\begin{table}
\centering
\caption{GSI and coherence for different levels of contact. Student's $t$-test $p$-values for the difference between the levels of contact are also shown.}
\label{tab:sync_stats_contact}
\begin{tabular}{lcccc}
\toprule
Contact state & \multicolumn{2}{c}{GSI} & \multicolumn{2}{c}{CWC} \\
\midrule
No contact & $0.15 \pm 0.05$ & \multirow{2}{*}{$2.93 \times 10^{-1}$} &$0.35 \pm 0.09$ & \multirow{2}{*}{$9.88 \times 10^{-2}$}  \\
Contact & $0.17 \pm 0.03$ & &$0.40 \pm 0.10$ & \\
\midrule
$B_r$ & $0.13 \pm 0.05$ & & $0.29 \pm 0.10$ & \\
$B_c$ & $0.14 \pm 0.05$ & & $0.30 \pm 0.09$ & \\
\bottomrule
\end{tabular}
\end{table}

%% file: tables/dunn_determinism.tex
\begin{table}
\centering
\caption{Dunn post-hoc test for pairwise comparisons of the determinism $D$ between baseline pairs of $B_r$ and $B_c$ as well as different intensities of interaction. The $p$-values are adjusted using the Bonferroni correction.}
\label{tab:dunn_determinism}
\begin{tabular}{lcccccc}
\toprule
 & $B_r$ & $B_c$ & 0 & 1 & 2 & 3 \\
\midrule
$B_r$ & - & $1.00$ & $8.10 \times 10^{-2}$ & $4.23 \times 10^{-1}$ & $\mathbf{< 10^{-4}}$ & $\mathbf{< 10^{-4}}$ \\
$B_c$ & - & - & $8.10 \times 10^{-2}$ & $4.23 \times 10^{-1}$ & $\mathbf{< 10^{-4}}$ & $\mathbf{< 10^{-4}}$ \\
0 & - & - & - & $1.00$ & $1.00$ & $3.20 \times 10^{-1}$ \\
1 & - & - & - & - & $4.13 \times 10^{-1}$ & $\mathbf{3.13 \times 10^{-2}}$ \\
2 & - & - & - & - & - & $3.10 \times 10^{-1}$ \\
3 & - & - & - & - & - & - \\
\bottomrule
\end{tabular}
\end{table}

%% file: tables/ssmd_determinism.tex
\begin{table}
\centering
\caption{SSMD for pairwise comparisons of the determinism $D$ between different intensities of interaction.}
\label{tab:ssmd_determinism}
\begin{tabular}{lcccccc}
\toprule
 & $B_r$ & $B_c$ & 0 & 1 & 2 & 3 \\
\midrule
$B_r$ & - & $-7.99 \times 10^{-3}$ & $-1.97 \times 10^{-1}$ & $-1.20 \times 10^{-1}$ & $-2.18 \times 10^{-1}$ & $-4.03 \times 10^{-1}$ \\
$B_c$ & - & - & $-1.97 \times 10^{-1}$ & $-1.17 \times 10^{-1}$ & $-2.19 \times 10^{-1}$ & $-4.15 \times 10^{-1}$ \\
0 & - & - & - & $8.53 \times 10^{-2}$ & $-2.46 \times 10^{-2}$ & $-2.42 \times 10^{-1}$ \\
1 & - & - & - & - & $-1.09 \times 10^{-1}$ & $-3.23 \times 10^{-1}$ \\
2 & - & - & - & - & - & $-2.17 \times 10^{-1}$ \\
3 & - & - & - & - & - & - \\
\bottomrule
\end{tabular}
\end{table}

%% file: tables/dunn_lyapunov.tex
\begin{table}
\centering
\caption{Dunn post-hoc test for pairwise comparisons of the maximal Lyapunov exponent $l_{lyap}$ between baseline pairs of $B_r$ and $B_c$ as well as different intensities of interaction. The $p$-values are adjusted using the Bonferroni correction.}
\label{tab:dunn_lyapunov}
\begin{tabular}{lcccccc}
\toprule
 & $B_r$ & $B_c$ & 0 & 1 & 2 & 3 \\
\midrule
$B_r$ & - & $3.44 \times 10^{-1}$ & $1.00$ & $1.75 \times 10^{-1}$ & $\mathbf{< 10^{-4}}$ & $\mathbf{< 10^{-4}}$ \\
$B_c$ & - & - & $7.82 \times 10^{-1}$ & $5.03 \times 10^{-1}$ & $\mathbf{2.12 \times 10^{-3}}$ & $\mathbf{< 10^{-4}}$ \\
0 & - & - & - & $3.44 \times 10^{-1}$ & $5.46 \times 10^{-2}$ & $\mathbf{< 10^{-4}}$ \\
1 & - & - & - & - & $1.00$ & $\mathbf{5.30 \times 10^{-3}}$ \\
2 & - & - & - & - & - & $\mathbf{2.82 \times 10^{-3}}$ \\
3 & - & - & - & - & - & - \\
\bottomrule
\end{tabular}
\end{table}

%% file: tables/ssmd_lyapunov.tex
\begin{table}
\centering
\caption{SSMD for pairwise comparisons of the maximal Lyapunov exponent $l_{lyap}$ between different intensities of interaction.}
\label{tab:ssmd_lyapunov}
\begin{tabular}{lcccccc}
\toprule
 & $B_r$ & $B_c$ & 0 & 1 & 2 & 3 \\
\midrule
$B_r$ & - & $6.79 \times 10^{-2}$ & $5.58 \times 10^{-2}$ & $2.06 \times 10^{-1}$ & $2.01 \times 10^{-1}$ & $4.10 \times 10^{-1}$ \\
$B_c$ & - & - & $-2.14 \times 10^{-2}$ & $1.64 \times 10^{-1}$ & $1.58 \times 10^{-1}$ & $4.28 \times 10^{-1}$ \\
0 & - & - & - & $2.20 \times 10^{-1}$ & $2.07 \times 10^{-1}$ & $5.42 \times 10^{-1}$ \\
1 & - & - & - & - & $4.26 \times 10^{-3}$ & $3.56 \times 10^{-1}$ \\
2 & - & - & - & - & - & $3.13 \times 10^{-1}$ \\
3 & - & - & - & - & - & - \\
\bottomrule
\end{tabular}
\end{table}

%% file: tables/dunn_rec.tex
\begin{table}
\centering
\caption{Dunn post-hoc test for pairwise comparisons of the percentage of recurrence $\%\text{REC}$ between baseline pairs of $B_r$ and $B_c$ as well as different intensities of interaction. The $p$-values are adjusted using the Bonferroni correction.}
\label{tab:dunn_rec}
\begin{tabular}{lcccccc}
\toprule
 & $B_r$ & $B_c$ & 0 & 1 & 2 & 3 \\
\midrule
$B_r$ & - & $\mathbf{< 10^{-4}}$ & $5.52 \times 10^{-1}$ & $\mathbf{1.19 \times 10^{-3}}$ & $\mathbf{< 10^{-4}}$ & $\mathbf{< 10^{-4}}$ \\
$B_c$ & - & - & $4.34 \times 10^{-1}$ & $9.16 \times 10^{-1}$ & $3.75 \times 10^{-1}$ & $\mathbf{5.51 \times 10^{-3}}$ \\
0 & - & - & - & $5.52 \times 10^{-1}$ & $1.03 \times 10^{-1}$ & $\mathbf{1.54 \times 10^{-3}}$ \\
1 & - & - & - & - & $9.16 \times 10^{-1}$ & $6.25 \times 10^{-2}$ \\
2 & - & - & - & - & - & $8.14 \times 10^{-2}$ \\
3 & - & - & - & - & - & - \\
\bottomrule
\end{tabular}
\end{table}

%% file: tables/ssmd_rec.tex
\begin{table}
\centering
\caption{SSMD for pairwise comparisons of the percentage of recurrence $\%\text{REC}$ between different intensities of interaction.}
\label{tab:ssmd_rec}
\begin{tabular}{lcccccc}
\toprule
 & $B_r$ & $B_c$ & 0 & 1 & 2 & 3 \\
\midrule
$B_r$ & - & $-3.02 \times 10^{-1}$ & $-1.50 \times 10^{-1}$ & $-3.30 \times 10^{-1}$ & $-3.92 \times 10^{-1}$ & $-7.07 \times 10^{-1}$ \\
$B_c$ & - & - & $1.70 \times 10^{-1}$ & $-1.22 \times 10^{-3}$ & $-7.43 \times 10^{-2}$ & $-3.74 \times 10^{-1}$ \\
0 & - & - & - & $-1.87 \times 10^{-1}$ & $-2.58 \times 10^{-1}$ & $-5.85 \times 10^{-1}$ \\
1 & - & - & - & - & $-7.95 \times 10^{-2}$ & $-4.06 \times 10^{-1}$ \\
2 & - & - & - & - & - & $-3.12 \times 10^{-1}$ \\
3 & - & - & - & - & - & - \\
\bottomrule
\end{tabular}
\end{table}

%% file: tables/dunn_det.tex
\begin{table}
\centering
\caption{Dunn post-hoc test for pairwise comparisons of the percentage of determinism $\%\text{DET}$ between baseline pairs of $B_r$ and $B_c$ as well as different intensities of interaction. The $p$-values are adjusted using the Bonferroni correction.}
\label{tab:dunn_det}
\begin{tabular}{lcccccc}
\toprule
 & $B_r$ & $B_c$ & 0 & 1 & 2 & 3 \\
\midrule
$B_r$ & - & $\mathbf{1.41 \times 10^{-4}}$ & $5.18 \times 10^{-1}$ & $\mathbf{8.39 \times 10^{-4}}$ & $\mathbf{< 10^{-4}}$ & $\mathbf{< 10^{-4}}$ \\
$B_c$ & - & - & $9.91 \times 10^{-1}$ & $1.38 \times 10^{-1}$ & $\mathbf{< 10^{-4}}$ & $\mathbf{< 10^{-4}}$ \\
0 & - & - & - & $5.18 \times 10^{-1}$ & $1.06 \times 10^{-1}$ & $\mathbf{2.76 \times 10^{-4}}$ \\
1 & - & - & - & - & $9.82 \times 10^{-1}$ & $\mathbf{1.42 \times 10^{-2}}$ \\
2 & - & - & - & - & - & $\mathbf{1.42 \times 10^{-2}}$ \\
3 & - & - & - & - & - & - \\
\bottomrule
\end{tabular}
\end{table}

%% file: tables/ssmd_det.tex
\begin{table}
\centering
\caption{SSMD for pairwise comparisons of the percentage of determinism $\%\text{DET}$ between different intensities of interaction.}
\label{tab:ssmd_det}
\begin{tabular}{lcccccc}
\toprule
 & $B_r$ & $B_c$ & 0 & 1 & 2 & 3 \\
\midrule
$B_r$ & - & $-6.38 \times 10^{-2}$ & $-2.53 \times 10^{-1}$ & $-3.25 \times 10^{-1}$ & $-3.27 \times 10^{-1}$ & $-4.68 \times 10^{-1}$ \\
$B_c$ & - & - & $-2.23 \times 10^{-1}$ & $-3.16 \times 10^{-1}$ & $-3.16 \times 10^{-1}$ & $-5.03 \times 10^{-1}$ \\
0 & - & - & - & $-1.69 \times 10^{-1}$ & $-1.69 \times 10^{-1}$ & $-5.74 \times 10^{-1}$ \\
1 & - & - & - & - & $-1.63 \times 10^{-2}$ & $-4.38 \times 10^{-1}$ \\
2 & - & - & - & - & - & $-3.58 \times 10^{-1}$ \\
3 & - & - & - & - & - & - \\
\bottomrule
\end{tabular}
\end{table}

%% file: tables/dunn_maxline.tex
\begin{table}
\centering
\caption{Dunn post-hoc test for pairwise comparisons of the maximal line length $\text{MAXLINE}$  between baseline pairs of $B_r$ and $B_c$ as well as different intensities of interaction. The $p$-values are adjusted using the Bonferroni correction.}
\label{tab:dunn_maxline}
\begin{tabular}{lcccccc}
\toprule
 & $B_r$ & $B_c$ & 0 & 1 & 2 & 3 \\
\midrule
$B_r$ & - & $\mathbf{< 10^{-4}}$ & $\mathbf{8.41 \times 10^{-3}}$ & $\mathbf{< 10^{-4}}$ & $\mathbf{< 10^{-4}}$ & $\mathbf{< 10^{-4}}$ \\
$B_c$ & - & - & $5.47 \times 10^{-1}$ & $\mathbf{3.56 \times 10^{-4}}$ & $\mathbf{< 10^{-4}}$ & $\mathbf{< 10^{-4}}$ \\
0 & - & - & - & $1.81 \times 10^{-1}$ & $\mathbf{8.41 \times 10^{-3}}$ & $\mathbf{< 10^{-4}}$ \\
1 & - & - & - & - & $5.47 \times 10^{-1}$ & $\mathbf{4.43 \times 10^{-3}}$ \\
2 & - & - & - & - & - & $\mathbf{8.41 \times 10^{-3}}$ \\
3 & - & - & - & - & - & - \\
\bottomrule
\end{tabular}
\end{table}

%% file: tables/ssmd_maxline.tex
\begin{table}
\centering
\caption{SSMD for pairwise comparisons of the maximal line length $\text{MAXLINE}$  between different intensities of interaction.}
\label{tab:ssmd_maxline}
\begin{tabular}{lcccccc}
\toprule
 & $B_r$ & $B_c$ & 0 & 1 & 2 & 3 \\
\midrule
$B_r$ & - & $-1.90 \times 10^{-1}$ & $-2.70 \times 10^{-1}$ & $-4.78 \times 10^{-1}$ & $-6.17 \times 10^{-1}$ & $-9.52 \times 10^{-1}$ \\
$B_c$ & - & - & $-6.57 \times 10^{-2}$ & $-3.12 \times 10^{-1}$ & $-4.50 \times 10^{-1}$ & $-8.21 \times 10^{-1}$ \\
0 & - & - & - & $-2.69 \times 10^{-1}$ & $-4.13 \times 10^{-1}$ & $-7.99 \times 10^{-1}$ \\
1 & - & - & - & - & $-1.25 \times 10^{-1}$ & $-5.42 \times 10^{-1}$ \\
2 & - & - & - & - & - & $-4.39 \times 10^{-1}$ \\
3 & - & - & - & - & - & - \\
\bottomrule
\end{tabular}
\end{table}